\documentclass[twocolumn,trackchanges]{aastex631}
\usepackage[utf8]{inputenc}
\usepackage{mathtools}
\usepackage{graphicx}
\usepackage{outlines}
\usepackage{color}
\usepackage{acronym}
\usepackage{xspace}
\let\tablenum\relax
\usepackage{siunitx}
\usepackage{placeins}

\setcitestyle{notesep={ }}

\definecolor{darkred}{rgb}{0.64, 0.0, 0.0} 

\newcommand{\Fstat}{$\F$-statistic\xspace}
\acrodef{gw}[GW]{gravitational-wave}
\acrodefplural{gw}[GWs]{gravitational waves}
\newcommand{\gw}{\ac{gw}\xspace}
\newcommand{\gws}{\acp{gw}\xspace}
\acrodef{cw}[CW]{continuous gravitational wave}
\acrodefplural{cw}[CWs]{Continuous gravitational waves} 
\newcommand{\cw}{\ac{cw}\xspace}
\newcommand{\cws}{\acp{cw}\xspace}
\acrodef{ns}[NS]{neutron star}
\acrodefplural{nss}[NSs]{neutron stars}
\newcommand{\ns}{\ac{ns}\xspace}
\newcommand{\nss}{\acp{ns}\xspace}
\acrodef{elmag}[EM]{electromagnetic}
\newcommand{\elmag}{\ac{elmag}\xspace}
\acrodef{snr}[S/N]{signal-to-noise ratio}
\acrodefplural{snrs}[S/N]{signal-to-noise ratios}
\newcommand{\snr}{\ac{snr}\xspace}

\acrodef{sft}[SFT]{short Fourier transform}
\acrodefplural{sft}[SFTs]{short Fourier transforms}
\newcommand{\sft}{\ac{sft}\xspace}
\newcommand{\sfts}{\acp{sft}\xspace}
\acrodef{fft}[FFT]{fast Fourier transform}
\acrodefplural{fft}[FFTs]{fast Fourier transforms}

\newcommand{\ffts}{\acp{fft}\xspace}
\acrodef{sfdb}[SFDB]{Short Fourier Database}
\acrodefplural{sfdb}[SFDBs]{Short Fourier Databases}

\newcommand{\sfdbs}{\acp{sfdb}\xspace}
\acrodef{psd}[PSD]{power spectral density}
\acrodefplural{psd}[PSDs]{power spectral densities}
\newcommand{\psd}{\ac{psd}\xspace}

\acrodef{ul}[UL]{upper limit}
\acrodefplural{uls}[UL]{upper limits}
\newcommand{\ul}{\ac{ul}\xspace}
\newcommand{\uls}{\acp{ul}\xspace}
\acrodef{ohthree}[O3]{third observing run}
\newcommand{\ohthree}[1][]{\ac{ohthree}#1\xspace}

\acrodef{most}[MOST]{Molonglo Observatory Synthesis Telescope}
\acrodef{chime}[CHIME]{Canadian Hydrogen Intensity Mapping Experiment}
\acrodef{nicer}[NICER]{Neutron Star Interior Composition Explorer}

\newcommand{\code}[1]{\texttt{#1}\xspace}
\newcommand{\lalsuite}{\code{LALSuite}}
\newcommand{\weave}{\code{lalapps\_Weave}}
\newcommand{\cfs}{\code{lalapps\_ComputeFstatistic\_v2}}
\newcommand{\pyfstat}{\code{PyFstat}}
\newcommand{\dmax}{\code{distromax}}

\newcommand{\tpsr}{t'} 
\newcommand{\tdet}{t} 
\newcommand{\Izz}{\mathcal{I}} 
\newcommand{\pdet}{p_\mathrm{det}} 
\newcommand{\hul}{h_0^{95\%}} 
\newcommand{\days}{\mathrm{days}}
\newcommand{\fspin}{f_\mathrm{spin}} 
\newcommand{\fdotspin}{\dot{f}_\mathrm{spin}} 
\newcommand{\fddotspin}{\ddot{f}_\mathrm{spin}} 
\newcommand{\fgw}{f} 
\newcommand{\fdot}{{\dot{\fgw}}} 
\newcommand{\fddot}{\ddot{\fgw}} 
\newcommand{\fdddot}{\dddot{\fgw}} 
\newcommand{\Tref}{T_\mathrm{ref}} 
\newcommand{\Amp}{\mathcal{A}} 
\newcommand{\Dop}{\lambda} 
\newcommand{\F}{\mathcal{F}} 
\newcommand{\Tsft}{T_\mathrm{SFT}} 
\newcommand{\TP}{\mathcal{T}} 
\newcommand{\Ntemp}{N_{\Dop}} 
\newcommand{\glitch}{\mathrm{gl}}
\newcommand{\Tgl}{T_\glitch} 
\newcommand{\win}{\varpi} 
\newcommand{\tzero}{\tdet_0} 
\newcommand{\Fmn}{\F_{mn}} 
\newcommand{\maxF}{\max2\F} 
\newcommand{\Btrans}{B_{\mathrm{tS}/\mathrm{G}}} 
\newcommand{\tCWdetstat}{\Btrans} 
\newcommand{\dtzero}{\mathrm{d}\tzero} 
\newcommand{\dtau}{\mathrm{d}\tau} 
\newcommand{\glstepf}{\Delta f_\glitch} 
\newcommand{\glstepfdot}{\Delta \fdot_\glitch} 
\newcommand{\glstepfddot}{\Delta \fddot_\glitch} 

\newcommand{\hsd}{h_\mathrm{sd}} 

\newcommand{\psr}[1]{PSR~J#1\xspace}
\newcommand{\crab}{\psr{0534+2200}}
\newcommand{\ohfivethirtyseven}{\psr{0537--6910}}
\newcommand{\ohnineoheight}{\psr{0908--4913}}
\newcommand{\elevenohfive}{\psr{1105--6107}}
\newcommand{\eighteenthirteen}{\psr{1813--1749}}
\newcommand{\eighteentwentysix}{\psr{1826--1334}}
\newcommand{\twentytwentyone}{\psr{2021+3651}}
\newcommand{\numglitchtargets}{six\xspace}
\newcommand{\numglitchtargetswithoutbigglitcher}{five\xspace}
\newcommand{\numglitchsearches}{nine\xspace}
\newcommand{\numglitchsearchesfordistromax}{seven\xspace}
\newcommand{\numglitchsearcheswithnooutliers}{eight\xspace}
\newcommand{\numnarrowbandtargets}{eighteen\xspace}
\newcommand{\numnarrowbandexceedspindwon}{seven\xspace}
\newcommand{\spdownfactor}{3\xspace}

\begin{document}
\title{Narrowband searches for continuous and long-duration transient gravitational waves from known pulsars in the LIGO-Virgo third observing run}


\author{R.~Abbott}
\affiliation{LIGO Laboratory, California Institute of Technology, Pasadena, CA 91125, USA}
\author{T.~D.~Abbott}
\affiliation{Louisiana State University, Baton Rouge, LA 70803, USA}
\author{F.~Acernese}
\affiliation{Dipartimento di Farmacia, Universit\`a di Salerno, I-84084 Fisciano, Salerno, Italy}
\affiliation{INFN, Sezione di Napoli, Complesso Universitario di Monte S. Angelo, I-80126 Napoli, Italy}
\author{K.~Ackley}
\affiliation{OzGrav, School of Physics \& Astronomy, Monash University, Clayton 3800, Victoria, Australia}
\author{C.~Adams}
\affiliation{LIGO Livingston Observatory, Livingston, LA 70754, USA}
\author{N.~Adhikari}
\affiliation{University of Wisconsin-Milwaukee, Milwaukee, WI 53201, USA}
\author{R.~X.~Adhikari}
\affiliation{LIGO Laboratory, California Institute of Technology, Pasadena, CA 91125, USA}
\author{V.~B.~Adya}
\affiliation{OzGrav, Australian National University, Canberra, Australian Capital Territory 0200, Australia}
\author{C.~Affeldt}
\affiliation{Max Planck Institute for Gravitational Physics (Albert Einstein Institute), D-30167 Hannover, Germany}
\affiliation{Leibniz Universit\"at Hannover, D-30167 Hannover, Germany}
\author{D.~Agarwal}
\affiliation{Inter-University Centre for Astronomy and Astrophysics, Pune 411007, India}
% ...

%


\collaboration{0}{The LIGO Scientific Collaboration, the Virgo Collaboration, and the KAGRA Collaboration}
\author{D.~Antonopoulou}
\affiliation{Jodrell Bank Centre for Astrophysics, School of Physics and Astronomy, University of Manchester, Manchester, UK, M13 9PL}

\author{Z.~Arzoumanian}
\affiliation{Astrophysics Science Division, NASA's Goddard Space Flight Center, Greenbelt, MD 20771, USA}

\author[0000-0002-4142-7831]{A.~Basu}
\affiliation{Jodrell Bank Centre for Astrophysics, School of Physics and Astronomy, University of Manchester, Manchester, UK, M13 9PL}

\author{S.~Bogdanov}
\affiliation{Columbia Astrophysics Laboratory, Columbia University, 550 West 120th Street, New York, NY, 10027, USA}

\author[0000-0002-1775-9692]{I.~Cognard}
\affiliation{Laboratoire de Physique et Chimie de l'Environnement et de l'Espace, Universit\'e d’Orl\'eans/CNRS, F-45071 Orl\'eans Cedex 02, France}
\affiliation{Station de radioastronomie de Nan\c{c}ay, Observatoire de Paris, CNRS/INSU, F-18330 Nan\c{c}ay, France}

\author[0000-0002-1529-5169]{K.~Crowter}
\affiliation{Department of Physics and Astronomy, University of British Columbia, 6224 Agricultural Road, Vancouver, BC V6T 1Z1 Canada}

\author{T.~Enoto}
\affiliation{RIKEN Cluster for Pioneering Research, 2-1 Hirosawa, Wako, Saitama 351-0198, Japan}

\author{C.~M. Espinoza}
\affiliation{Departamento de F\'isica, Universidad de Santiago de Chile, 9170124 Estaci\'on Central, Chile}
\affiliation{Center for Interdisciplinary Research in Astrophysics and Space Sciences, Universidad de Santiago de Chile, Santiago, Chile}

\author[0000-0003-1110-0712]{C.~M.~L.~Flynn}
\affiliation{OzGrav, Swinburne University of Technology, Hawthorn VIC 3122, Australia}

\author[0000-0001-8384-5049]{E.~Fonseca}
\affiliation{Department of Physics and Astronomy, West Virginia University, PO Box 6315, Morgantown, WV 26506, USA} 
\affiliation{Center for Gravitational Waves and Cosmology, West Virginia University, Chestnut Ridge Research Building, Morgantown, WV 26505, USA}

\author[0000-0003-1884-348X]{D.~C.~Good}
\affiliation{Center for Computational Astrophysics, Flatiron Institute, 162 5th Avenue, New York, New York, 10010, USA}
\affiliation{Department of Physics, University of Connecticut, 196 Auditorium Road, U-3046, Storrs, CT 06269-3046, USA}

\author[0000-0002-9049-8716]{L.~Guillemot}
\affiliation{Laboratoire de Physique et Chimie de l'Environnement et de l'Espace, Universit\'e d’Orl\'eans/CNRS, F-45071 Orl\'eans Cedex 02, France}
\affiliation{Station de radioastronomie de Nan\c{c}ay, Observatoire de Paris, CNRS/INSU, F-18330 Nan\c{c}ay, France}

\author[0000-0002-6449-106X]{S.~Guillot}
\affiliation{IRAP, CNRS, 9 avenue du Colonel Roche, BP 44346, F-31028 Toulouse Cedex 4, France}
\affiliation{Universit\'e de Toulouse, CNES, UPS-OMP, F-31028 Toulouse, France}

\author{A.~K.~Harding}
\affiliation{Los Alamos National Laboratory, Los Alamos, NM 87545, USA}

\author[0000-0001-5567-5492]{M.~J.~Keith}
\affiliation{Jodrell Bank Centre for Astrophysics, School of Physics and Astronomy, University of Manchester, Manchester, UK, M13 9PL}

\author{L.~Kuiper}
\affiliation{SRON-Netherlands Institute for Space Research, Niels Bohrweg 4, 2333 CA, Leiden, Netherlands}

\author{M.~E.~Lower}
\affiliation{Centre for Astrophysics and Supercomputing, Swinburne University of Technology, Hawthorn, VIC, 3122, Australia}
\affiliation{CSIRO, Space and Astronomy, PO Box 76, Epping, NSW, 1710, Australia}

\author{A.~G.~Lyne}
\affiliation{Jodrell Bank Centre for Astrophysics, School of Physics and Astronomy, University of Manchester, Manchester, UK, M13 9PL}

\author[0000-0002-2885-8485]{J.~W.~McKee}
\affiliation{Canadian Institute for Theoretical Astrophysics, University of Toronto, 60 St. George Street, Toronto, ON M5S 3H8, Canada}

\author[0000-0001-8845-1225]{B.~W.~Meyers}
\affiliation{Department of Physics and Astronomy, University of British Columbia, 6224 Agricultural Road, Vancouver, BC V6T 1Z1 Canada}

\author{C.~Ng}
\affiliation{Dunlap Institute for Astronomy \& Astrophysics, University of Toronto, 50 St.~George Street, Toronto, ON M5S 3H4, Canada}

\author{J.~L.~Palfreyman}
\affiliation{School of Natural Sciences, University of Tasmania, Hobart, Australia}

\author[0000-0002-7285-6348]{R.~M.~Shannon}
\affiliation{Centre for Astrophysics and Supercomputing, Swinburne University of Technology, Hawthorn, VIC, 3122, Australia}
\affiliation{ARC Centre of Excellence for Gravitational Wave Discovery (OzGrav)}

\author[0000-0002-9581-2452]{B.~Shaw}
\affiliation{Jodrell Bank Centre for Astrophysics, School of Physics and Astronomy, University of Manchester, Manchester, UK, M13 9PL}

\author[0000-0001-9784-8670]{I.~H.~Stairs}
\affiliation{Department of Physics and Astronomy, University of British Columbia, 6224 Agricultural Road, Vancouver, BC V6T 1Z1 Canada}

\author[0000-0001-9242-7041]{B.~W.~Stappers}
\affiliation{Jodrell Bank Centre for Astrophysics, School of Physics and Astronomy, University of Manchester, Manchester, UK, M13 9PL}

\author[0000-0001-7509-0117]{C.~M.~Tan}
\affiliation{Department of Physics, McGill University, 3600 rue University, Montr\'{e}al, QC H3A 2T8, Canada}                                       
\affiliation{McGill Space Institute, McGill University, 3550 rue University, Montr\'{e}al, QC H3A 2A7, Canada}

\author[0000-0002-3649-276X]{G.~Theureau}
\affiliation{Laboratoire de Physique et Chimie de l'Environnement et de l'Espace, Universit\'e d’Orl\'eans/CNRS, F-45071 Orl\'eans Cedex 02, France}
\affiliation{Station de radioastronomie de Nan\c{c}ay, Observatoire de Paris, CNRS/INSU, F-18330 Nan\c{c}ay, France}
\affiliation{LUTH, Observatoire de Paris, PSL Research University, CNRS, Universit\'e Paris Diderot, Sorbonne Paris Cit\'e, F-92195 Meudon, France}

\author[0000-0003-2122-4540]{P.~Weltevrede}
\affiliation{Jodrell Bank Centre for Astrophysics, School of Physics and Astronomy, University of Manchester, Manchester, UK, M13 9PL}

\vspace*{-2\baselineskip}
\date{\today}

\begin{abstract}
     Isolated neutron stars that are asymmetric with respect to their spin axis are possible sources of detectable continuous gravitational waves. This paper presents a fully-coherent search for such signals from \numnarrowbandtargets pulsars in data from LIGO and Virgo's third observing run (O3). For known pulsars, efficient and sensitive matched-filter searches can be carried out if one assumes the gravitational radiation is phase-locked to the electromagnetic emission. In the search presented here, we relax this assumption and allow the frequency and frequency time-derivative of the gravitational waves to vary in a small range around those inferred from electromagnetic observations. We find no evidence for continuous gravitational waves, and set upper limits on the strain amplitude for each target. These limits are more constraining for \numnarrowbandexceedspindwon of the targets than the spin-down limit defined by ascribing all rotational energy loss to gravitational radiation. In an additional search we look in O3 data for long-duration (hours--months) transient gravitational waves in the aftermath of pulsar glitches for \numglitchtargets targets with a total of \numglitchsearches glitches. We report two marginal outliers from this search, but find no clear evidence for such emission either. The resulting duration-dependent strain upper limits do not surpass indirect energy constraints for any of these targets.
\end{abstract}

\section{Introduction}
\begin{outline}

\cws are quasi-monochromatic signals expected to be ever-present in the data
of \gw detectors such as Advanced LIGO~\citep{TheLIGOScientific:2014jea} and Advanced Virgo~\citep{2015CQGra..32b4001A}.
While the observation of transient \gws from compact binary coalescences has become nearly commonplace~\citep{LIGOScientific:2020ibl},
\cws have yet to be detected as of the \ohthree.
One of the most enticing and commonly sought after sources of \cws is a rapidly spinning, asymmetric \ns; see \citet{Sieniawska:2019hmd} and \citet{Haskell:2021ljd} for recent reviews.
In the case of a triaxial \ns, \cw emission occurs at twice the rotation frequency of the star.

Many \nss are observed as pulsars by radio, X-ray or $\gamma$-ray telescopes~\citep{LorimerKramer2012}.
Pulsars can often be timed extremely precisely---in the best cases the arrival of new pulses can be predicted to within tens of nano-seconds.
This precision can enable exciting science, including sensitive tests of general relativity~\citep{Wex:2020ald},
placing constraints on the equation of state of the dense matter inside \nss~\citep{Lattimer:2004pg,KramerWex2009,HoEspinoza2015},
and using deviations in timing residuals of pulsars to search for a stochastic gravitational-wave background~\citep{VerbiestLentati2016,ArzoumanianBaker2020,GoncharovShannon2021}.
Detecting \gws from spinning \nss would add a completely new messenger to the study of these extreme objects~\citep{Glampedakis:2017nqy,Haskell:2021ljd}.

In this paper, we search in LIGO--Virgo \ohthree data
(taken in 2019--2020)
for \cws from \nss that have been observed as pulsars and precisely timed in either the radio or X-ray bands.
We identify \numnarrowbandtargets promising candidates for which the observed spin-down
(negative frequency derivative)
implies indirect limits on \cw emission
that fall within a factor of \spdownfactor of the expected sensitivity of the search.
Some other analyses assume the phase of \cws to be locked to the rotational phase of the crust of the star as observed by \elmag telescopes
\citep{2014ApJ...785..119A,2017ApJ...839...12A,2018PhRvL.120c1104A,2019ApJ...879...10A,LIGOScientific:2020gml,LIGOScientific:2020lkw,O3KnownPulsars,AshokBeheshtipour2021,Nieder:2020yqy}.
Here we relax that assumption, and allow for the frequency of rotation, and its derivative, to differ from the \elmag-observed values by a small factor:
the so-called ``narrowband'' search approach
\citep{2008ApJ...683L..45A,2015PhRvD..91b2004A,2017PhRvD..96l2006A,O2Narrowband, AshokBeheshtipour2021,Nieder:2020yqy}.
We use two separate analysis pipelines, the $5n$-vector~\citep{AstoneColla2014,2017CQGra..34m5007M} and the frequency-domain \Fstat~\citep{1998PhRvD..58f3001J,WettEtAl2018:ImpSmSCnGrvWUOCnTB} pipelines, to perform phase-coherent searches for \cws on \ohthree data over our widened parameter space. Using two separate pipelines allows us to cross-check results between pipelines, compare limits set by the two methods, and increase confidence in any potential detection by requiring both pipelines to see the same signal.

A scenario where \gw and \elmag emission have similar, but slightly different phase evolution is plausible,
\textit{e.g.}, when there is differential rotation between the rigid crust and superfluid parts of the star.
Possible observational evidence for this comes, for example, from pulsar glitches~\citep{Link:1992mdl,Link:2000mu,Lyne:2000sta,Fuentes:2017bjx,Haskell:2017ngx}:
sudden spin-up events that are often followed by an exponential relaxation back to the simple spin-down scenario.
In many models~\citep{Haskell:2015jra},
the superfluid and non-superfluid components build up a lag with respect to one another,
and a glitch represents the sudden re-coupling of the two components~\citep{1975Natur.256...25A,Lyne:2000sta}.

Glitches are also directly relevant for \gw searches:
first, some of our \cw search targets glitched during \ohthree.
For these, we perform separate phase-coherent searches covering the parts of \ohthree before and after the glitches.
Second, it is also possible that glitches trigger increased \gw emission in their aftermath~\citep{vanEysden:2008pd,Bennett:2010tm,Prix:2011qv,Melatos:2015oca,Singh:2016ilt,Yim:2020trr}.
Hence, we also perform additional searches for long-duration transient \cw-like signals
with durations from a few hours to four months
after observed glitches during (or shortly before) \ohthree,
covering \numglitchsearches glitches
from \numglitchtargets pulsars.

The rest of this paper is structured as follows.
We discuss our selection of target pulsars,
and the \elmag observations that we use to guide our search
in Sec. \ref{sec:targets}
and the \ohthree \gw~data set in Sec. \ref{sec:data}.
In Sec. \ref{sec:model-methods} we describe our analysis methods.
We present the results of the two \cw searches in Sec. \ref{sec:results}.
We then cover the detailed setup and the results of the search for \gw~emission in the aftermath of pulsar glitches in Sec.~\ref{sec:transient}.
In Sec.~\ref{sec:conclusion} we conclude the paper with a discussion of the results obtained by all three analyses and their astrophysical implications.
Throughout the rest of the paper we will often refer to the analyses using the \Fstat and 5$n$-vector pipelines search over all data as the two ``CW'' searches, and the search for long-duration transients after glitches as the ``transient search.''
\end{outline}

\section{Electromagnetic Data and Target Selection}
\label{sec:targets}
We start with timing solutions from \elmag observations of pulsars (also referred to as \emph{ephemerides}), and search in a narrow band in frequency and spin-down, as further described below in Sec.~\ref{sec:model-methods}.
The widths of the frequency and spin-down search bands are usually larger than the uncertainty in the timing solutions.
The observations we use were made in radio and X-ray bands,
and were provided by the following observatories:
the \ac{chime}~\citep{CHIME_reference,Amiri_2021},
University of Tasmania's Mount Pleasant Observatory 26~m telescope,
the 42~ft telescope and Lovell telescope at Jodrell Bank,
the MeerKAT telescope \citep[observations made as part of the MeerTime project,][]{MeerKAT_reference},
the Nan\c{c}ay Decimetric Radio Telescope,
the \ac{nicer}~\citep{NICER_reference}
and the UTMOST timing program with the \ac{most} ~\citep{UTMOST_reference,Jankowski2018abc,Lower:2020mjq}.
The \code{Tempo}~\citep{NiceDemorest2015} and \code{Tempo2}~\citep{Edwards:2006zg} timing packages were used to fit the model parameters and provide timing solutions.

We select our targets for the \cw searches in a similar manner to~\citet{O2Narrowband},
based on the sensitive frequency band of the Advanced LIGO and Virgo detectors
and availability of precise ephemerides over the duration of \ohthree from \elmag observations.
We have analyzed all isolated pulsars, except for one, with a rotation frequency between 10 and 350\,Hz and for which the spin-down limit falls within a factor of \spdownfactor of the expected sensitivity of the full network over the course of \ohthree.
This frequency range includes all the high-value targets identified in \citet{O3KnownPulsars} for which it could be possible to go below the spin-down limit.
Pulsars in this frequency range would produce \cws between 20~Hz and 700~Hz.
The only pulsar we do not analyze that satisfies these criteria is \ohfivethirtyseven, which was analyzed in detail using a narrowband approach to search for $r$-mode emission in~\citet{LIGOScientific:2021yby} and using a targeted approach in~\citet{LIGOScientific:2020lkw}.
Searches lasting up to 120~days during inter-glitch periods for this pulsar are performed by the post-glitch transient search presented in Sections~\ref{sec:transient-method} and~\ref{sec:transient-setup}.

We estimate the expected sensitivity as 
\begin{equation}
 \label{eq:sens-estimate}
 h_{\rm sens}=\Theta \sqrt{\frac{S(f)}{T_{\rm obs}}},
\end{equation}
where $S(f)$ is the power spectral density as a function of frequency for the LIGO Hanford, LIGO Livingston and Virgo detectors (H1, L1 and V1 respectively) and $T_{\rm obs}$ is the observing time assuming 11 months of data with a duty cycles of 75\%, 77\% and 76\% respectively.
The factor $\Theta \sim 30$ encodes the scaling of typical narrowband searches but is a function of the number of templates employed in the analysis \citep{AstoneColla2014}.

On the other hand, the spin-down limit is an indirect \ul on the \gw amplitude assuming all of a pulsar's rotational energy loss comes in the form of \gws~\citep{1998PhRvD..58f3001J,Prix:2009oha}. 
It is given by
\begin{equation}
 \label{eq:spindown-limit}
 \hsd \approx 2.55\times 10^{-25} \frac{1~\textrm{kpc}}{d} \sqrt{I_{38} \frac{|\fdotspin|}{10^{-13}\,\rm{Hz/s}} \frac{1\,\rm{Hz}}{\fspin} },
\end{equation}
where $I_{38}$ is the star's moment of inertia in units of $10^{38}~\rm{kg~m^2}$,
$\fspin$ is its rotation frequency,
and $|\fdotspin|$ is the absolute value of its spin-down rate~\citep{O2Narrowband}.
We have computed spin-down limits according to the most recent distance estimates given in the ATNF catalog~\citep{2005AJ....129.1993M}, version 1.64,
and extrapolating the rotational frequencies and spin-down rates to \ohthree. For several pulsars, we have used more recent distance estimates from the literature.

In Table~\ref{tab:search_setup_parameter} we present a list of targets
along with the ranges of \gw frequency and spin-down parameters
and corresponding number of templates
covered for the two pipelines used to conduct the \cw searches.
We discuss our parameter range choices for each pipeline in Sec.~\ref{sec:model-methods}.
The observatory yielding the timing solution we use for each source is noted in the far right column.

Two pulsars on our list of \cw search targets, \crab (the Crab) and \psr{1105--6107},
glitched during \ohthree~\citep{Shaw:2021vvs}, on 2019 July 23 and 2019 April 9 respectively.
One other, \eighteenthirteen, shows marginal evidence for a glitch on or close to 2019 Aug 03~\citep{Ho:2020oel}.

For this reason, we perform two separate searches, before and after the estimated glitch epoch,
which are identified in Table~\ref{tab:search_setup_parameter} with suffixes ``bg'' and ``ag'' for ``before glitch'' and ``after glitch'' respectively.

For \crab, we use O3 data until the last observation before the glitch.
The analysis after the glitch uses data from ten days after the glitch epoch, accounting for the estimated relaxation time, until the end of O3.

For \elevenohfive, we only search after the glitch since the glitch occurs nearly at the start of the run.
The after-glitch search uses data from two days after the estimated glitch until the end of O3. 
For \eighteenthirteen, we perform two separate searches:
one fully-coherent search across the full \ohthree duration, assuming the pulsar did not glitch,
and a search before the glitch. No search after the glitch is performed since the glitch occurs near the end of the \elmag timing model
(although a post-glitch transient search is conducted assuming the timing model remains valid within uncertainties; see below).
Further details about these glitches will also be discussed in Sec.~\ref{sec:transient}.

For the post-glitch transient search,
we have used information from the glitch catalogs maintained at ATNF~\citep{ATNFGlitchDatabase} and Jodrell Bank~\citep{2011MNRAS.414.1679E,Basu:2021pyd,JodrellGlitchCatalogue}.
We have selected \numglitchtargets pulsars
with \gw frequencies $\fgw>15$\,Hz
and with glitches observed during (or shortly before) O3:
\crab,
\ohfivethirtyseven,
\ohnineoheight,
\elevenohfive,
\eighteenthirteen,
and \eighteentwentysix.
Another pulsar within our frequency band of interest, \twentytwentyone,
was observed to glitch during O3 by Jodrell Bank,
but the glitch time uncertainty is too large to make our transient search setup feasible.
As mentioned before, for \eighteenthirteen it is not certain if a glitch actually occurred~\citep{Ho:2020oel},
but we perform an opportunistic search here with its assumed parameters.

The ephemerides for \crab
and \eighteentwentysix
were provided by Jodrell Bank;
a detailed discussion of the \crab glitch
is given in \citet{Shaw:2021vvs}.
Ephemerides for \ohnineoheight and \elevenohfive are derived from UTMOST radio observations;
these are discussed further in \citet{Lower:2019awx,Lower:2020mjq}.
A potential ambiguity in timing solutions from periodically-scheduled surveys like UTMOST,
particularly affecting glitch size estimates,
was discussed by~\citet{Dunn:2021ltx},
but for these two targets it has been confirmed that the provided values are correct.
For \eighteenthirteen and \ohfivethirtyseven we use \ac{nicer} X-ray observations,
as reported in \citet{Ho:2020oel,Ho:2020vxt} and \citet{LIGOScientific:2020lkw}.
Details for all targets will be listed in Sec.~\ref{sec:transient-setup}.

\section{GW data}
\label{sec:data}

We analyze \gw strain data taken at the LIGO Hanford Observatory (H1), LIGO Livingston Observatory (L1) and Virgo (V1), during their \ohthree observing run.
\ohthree consisted of two separate sections, separated by a month-long commissioning break.
\ohthree[a] ran from 2019 April 1, 15:00 UTC until 2019 October 1, 15:00 UTC.
\ohthree[b] ran from 2019 November 1, 15:00 UTC, to 2020 March 27, 17:00 UTC.
See \citet{aLIGO:2020wna} and \citet{PhysRevLett.123.231108} for general descriptions of the performances of the LIGO and Virgo detectors respectively during \ohthree.
The calibration of this data set and its uncertainty budget are described in \citet{Sun:2020wke,Sun:2021qcg} and \citet{VIRGO:2021kfv}.

Data preparation for all searches starts with removing times of large transient noise whose cause is known, referred to as ``CAT1'' vetoes~\citep{LIGO:2021ppb,Acernese:2022ozw} from calibrated strain data.
\movetabledown=5cm
\begin{rotatetable*}
\begin{deluxetable*}{lllccccccccccc}
\tablecaption{
\label{tab:search_setup_parameter}
Set-up parameters for fully coherent \cw search pipelines.
}
\tablehead{\colhead{Name} & \colhead{R.A.} & \colhead{DEC.} & \colhead{$\fgw$} & \colhead{$\Delta\fgw$ (5v)} & \colhead{$\Delta\fgw\,(\F)$} & \colhead{$\fdot$} & \colhead{$\Delta\fdot$ (5v)} &
\colhead{$\Delta\fdot\,(\F)$} &
\colhead{$\fddot$} &
\colhead{$\Delta\fddot$}&
\colhead{$n_{\rm total}^{\rm{5v}}$} & \colhead{$n_{\rm total}^{\F}$} & Ref. \\
\colhead{} & \colhead{} & \colhead{} & \colhead{} & \colhead{} & \colhead{} & \colhead{$\times 10^{-13}$} & \colhead{$\times 10^{-15}$} & \colhead{$\times 10^{-15}$} & \colhead{$\times 10^{-23}$} & \colhead{$\times 10^{-23}$} & \colhead{$\times 10^{7}$} & \colhead{$\times 10^7$} &\\
\colhead{} & \colhead{} & \colhead{} & \colhead{Hz} & \colhead{Hz} & \colhead{Hz} & \colhead{$\rm{Hz~s^{-1}}$} & \colhead{$\rm{Hz~s^{-1}}$} & \colhead{$\rm{Hz~s^{-1}}$} & \colhead{$\rm{Hz~s^{-2}}$} & \colhead{$\rm{Hz~s^{-2}}$} & \colhead{} & \colhead{} &
}
\startdata
   J0534+2200\textrm{ bg} &$ 05^{\textrm{h}}34^{\textrm{m}}31.97^{\textrm{s}} $&$  +22^\circ00'52.07'' $&$   59.241 $&$ \ldots $&$  0.24 $&$  -7370.0 $&$    \ldots $&$   2900.0 $&$ 2360.0 $&$     5.2 $&$     \ldots$ &$     983.7 $&a\\
   J0534+2200\textrm{ ag} &$ 05^{\textrm{h}}34^{\textrm{m}}31.97^{\textrm{s}} $&$  +22^\circ00'52.07'' $&$   59.241 $&$   0.24 $&$  0.24 $&$  -7370.0 $&$    2300.0 $&$   2900.0 $&$ 2360.0 $&$     5.2 $&$      498.0$ &$    9362.0 $    &a\\
               J0711--6830 &$ 07^{\textrm{h}}11^{\textrm{m}}54.18^{\textrm{s}} $&$ -68^\circ30'47.37''  $&$  364.234 $&$   0.72 $&$   1.5 $&$ -0.00989 $&$       2.1 $&$    0.004 $&$    0.0 $&$     0.0 $&$      4.391$ &$     45.85 $   &b\\
               J0835--4510 &$ 08^{\textrm{h}}35^{\textrm{m}}20.52^{\textrm{s}} $&$ -45^\circ10'34.28''  $&$   22.371 $&$  0.089 $&$ 0.089 $&$   -313.0 $&$     130.0 $&$    120.0 $&$  504.0 $&$     0.0 $&$       32.9$ &$     281.6 $   &c\\
               J1101--6101 &$ 11^{\textrm{h}}01^{\textrm{m}}44.96^{\textrm{s}} $&$  -61^\circ01'39.6''  $&$   31.846 $&$   0.13 $&$  0.13 $&$    -45.3 $&$      19.0 $&$     18.0 $&$    0.0 $&$     0.0 $&$      7.026$ &$     61.52 $   &d\\
               J1105--6107 &$ 11^{\textrm{h}}05^{\textrm{m}}25.71^{\textrm{s}} $&$ -61^\circ07'55.63'' $&$   31.644 $&$   0.13 $&$  0.13 $&$    -79.7 $&$      32.0 $&$     32.0 $&$  554.0 $&$    35.0 $&$      11.64 $&$     266.0 $   & e \\
               J1809--1917 &$ 18^{\textrm{h}}09^{\textrm{m}}43.13^{\textrm{s}} $&$  -19^\circ17'38.2''  $&$   24.166 $&$  0.097 $&$ 0.097 $&$    -74.4 $&$      32.0 $&$     30.0 $&$    3.7 $&$    0.14 $&$      8.886$ &$     152.3 $   &a\\
   J1813-1749\textrm{ bg} &$ 18^{\textrm{h}}13^{\textrm{m}}35.11^{\textrm{s}} $&$ -17^\circ49'57.57''  $&$   44.703 $&$ \ldots $&$  0.18 $&$  -1290.0 $&$    \ldots $&$    510.0 $&$    0.0 $&$     0.0 $&$     \ldots$ &$     92.08 $ &d\\
 J1813-1749\textrm{ full} &$ 18^{\textrm{h}}13^{\textrm{m}}35.11^{\textrm{s}} $&$ -17^\circ49'57.57''  $&$   44.703 $&$   0.18 $&$  0.18 $&$  -1290.0 $&$     520.0 $&$    510.0 $&$    0.0 $&$     0.0 $&$      266.4$ &$    2270.0 $   &d\\
               J1828--1101 &$ 18^{\textrm{h}}28^{\textrm{m}}18.85^{\textrm{s}} $&$ -11^\circ01'51.72''  $&$   27.754 $&$   0.11 $&$  0.11 $&$    -57.0 $&$      23.0 $&$     23.0 $&$   4.23 $&$     9.4 $&$      7.481$ &$    1162.0 $   &a\\
               J1833--0827 &$ 18^{\textrm{h}}33^{\textrm{m}}40.26^{\textrm{s}} $&$ -08^\circ27'31.53''  $&$   23.449 $&$  0.094 $&$ 0.094 $&$    -25.2 $&$      11.0 $&$     10.0 $&$ -0.463 $&$   0.082 $&$      2.873$ &$     242.2 $   &e\\
               J1838--0655 &$  18^{\textrm{h}}38^{\textrm{m}}3.13^{\textrm{s}} $&$  -06^\circ55'33.4''  $&$   28.363 $&$   0.11 $&$  0.11 $&$   -199.0 $&$      83.0 $&$     80.0 $&$   67.1 $&$     7.8 $&$      27.13$ &$     555.4 $   &d\\
               J1856+0245 &$ 18^{\textrm{h}}56^{\textrm{m}}50.91^{\textrm{s}} $&$ +02^\circ45'53.17''  $&$   24.714 $&$  0.099 $&$\ldots $&$   -189.0 $&$      79.0 $&$   \ldots $&$    0.0 $&$     0.0 $&$      22.42$ &$    \ldots $   &a\\
               J1913+1011 &$ 19^{\textrm{h}}13^{\textrm{m}}20.34^{\textrm{s}} $&$ +10^\circ11'22.97''  $&$   55.694 $&$   0.22 $&$  0.22 $&$    -52.5 $&$      23.0 $&$     21.0 $&$ -0.321 $&$     2.6 $&$      15.02$ &$    1951.0 $   &a\\
               J1925+1720 &$ 19^{\textrm{h}}25^{\textrm{m}}27.06^{\textrm{s}} $&$ +17^\circ20'27.42''  $&$   26.434 $&$   0.11 $&$  0.11 $&$    -36.6 $&$      15.0 $&$     15.0 $&$   3.93 $&$     6.3 $&$      4.538$ &$     469.2 $   &a\\
               J1928+1746 &$ 19^{\textrm{h}}28^{\textrm{m}}42.55^{\textrm{s}} $&$ +17^\circ46'29.67''  $&$   29.097 $&$   0.12 $&$  0.12 $&$    -55.7 $&$      23.0 $&$     22.0 $&$    0.0 $&$     0.0 $&$      7.845$ &$     68.41 $   &a\\
               J1935+2025 &$ 19^{\textrm{h}}35^{\textrm{m}}41.94^{\textrm{s}} $&$  +20^\circ25'40.1''  $&$   24.955 $&$  0.092 $&$   0.1 $&$   -189.0 $&$      79.0 $&$     76.0 $&$   95.0 $&$     3.4 $&$      20.99$ &$     581.3 $   &a\\
               J1952+3252 &$ 19^{\textrm{h}}52^{\textrm{m}}58.21^{\textrm{s}} $&$ +32^\circ52'40.51''  $&$   50.587 $&$    0.2 $&$   0.2 $&$    -74.9 $&$      32.0 $&$     30.0 $&$   2.92 $&$   0.012 $&$      18.61$ &$    3908.0 $   &f\\
               J2124--3358 &$ 21^{\textrm{h}}24^{\textrm{m}}43.84^{\textrm{s}} $&$ -33^\circ58'45.06''  $&$  405.588 $&$   0.91 $&$   1.6 $&$  -0.0169 $&$       2.1 $&$   0.0068 $&$    0.0 $&$     0.0 $&$      5.595$ &$     48.06 $   &g\\
               J2229+6114 &$  22^{\textrm{h}}29^{\textrm{m}}6.57^{\textrm{s}} $&$  +61^\circ14'10.9''  $&$   38.709 $&$   0.15 $&$  0.16 $&$   -590.0 $&$     240.0 $&$    260.0 $&$ 1170.0 $&$     0.0 $&$      107.3$ &$    1021.0 $   &f
\enddata
\tablecomments{
 This table lists frequency $\fgw$ and spin-down ($\fdot$,$\fddot$) parameters corresponding to twice
 the rotational parameters in the pulsar ephemerides at the start of the O3 run
 (2019 April 1 15:00 UTC, MJD 58574.000),
 as discussed in Sec.~\ref{sec:signal}.
 Right ascension (RA) and Declination (DEC) are given in J2000 coordinates.
 In the column headings,
 $\F$ stands for the $\F$-statistic search discussed in Sec.~\ref{sec:fstat-method}
 and $5$v stands for the 5$n$-vector method discussed in Sec.~\ref{sec:fivevec-method}.
 The suffixes ``bg'' and ``ag'' refer to ``before'' and ``after'' glitch respectively, while ``full'' corresponds to the full run in the case of \psr{1813--1749}. \\
(a) Jodrell Bank 42ft telescope and Lovell telescope
(b) MeerKAT telescope (part of the MeerTime project) \citep{MeerKAT_reference}
(c) University of Tasmania's Mt. Pleasant Observatory 26m telescope
(d) \ac{nicer}~\citep{NICER_reference}
(e) \ac{most}~\citep{UTMOST_reference,Jankowski2018abc,Lower:2020mjq}
(f) \ac{chime}~\citep[part of the \ac{chime} pulsar project,][]{CHIME_reference,Amiri_2021}
(g) Nan\c{c}ay Decimetric Radio Telescope.
This table is available online in a machine-readable format~\citep{data_release}.}
\end{deluxetable*}
\end{rotatetable*}
\noindent This procedure removed \SI{0.7}{\days} out of \SI{246}{days} of H1 data, \SI{2.2}{\days} out of \SI{256}{\days} of L1 data, and \SI{0.46}{\days} out of \SI{251}{\days} of V1 data.
 
The data are then cleaned to remove some known monochromatic or quasi-monochromatic detector artifacts that can be effectively monitored by external sensors, like signals injected for calibration, or contamination from power mains~\citep{Davis:2018yrz,T2100058,Sun:2020wke,Sun:2021qcg,VIRGO:2021kfv,Acernese:2022ozw}.
The H1 and L1 data near the 60\,Hz power mains frequency are also cleaned using a non-linear filtering method outlined in~\citet{VajenteHuang2020}.

The data from L1 and H1 are then ``gated'' to remove further large transient artifacts, which removes an additional \SI{1.0}{\days} of data from L1 and \SI{0.62}{\days} of data from H1~\citep{T2000384,LIGO:2021ppb}.

There are some differences in how each search pipeline uses these data, as will also be discussed in detail in Sec.~\ref{sec:model-methods}.
The 5$n$-vector search
creates \sfdbs from the time domain data.
\sfdbs consist of a collection of short-duration (\SI{1024}{s} long) \ffts, overlapped by half and Tukey-windowed with a window shape parameter of $\alpha=0.001$.
During the construction of the \sfdbs, an extra time-domain cleaning~\citep[][, in addition to those discussed in the previous paragraph]{Astone_2005}
is applied to identify delta-like time-domain disturbances.
The \Fstat and transient pipelines 
create \SI{1800}{\second} Tukey-windowed \sfts  with a window shape parameter of $\alpha=0.001$, and then extract a narrow frequency range
around the frequency band for each pulsar.

Additional narrow lines with identified instrumental origin are listed in \citet{linelist01} and \citet{virgo3lines};
we use these for pipeline-level cleaning or post-processing vetoes, as discussed in later sections.

\section{Signal model and search methods}
\label{sec:model-methods}

\subsection{Signal model}
\label{sec:signal}

With all three pipelines,
we search for quasi-monochromatic \gw signals
with fixed intrinsic amplitude
at approximately twice the pulsar rotation frequency,
\mbox{$\fgw \approx 2\fspin$},
allowing for some deviation in the narrowband search approach.
The general frequency evolution model for such \gws,
following the standard model for a pulsar's $\fspin$
as a function of time $\tpsr$ at the pulsar,
is
\begin{equation}
\label{eq:frequency-model}
f(\tpsr) = f_\mathrm{Taylor}(\tpsr) + f_\mathrm{glitch}(\tpsr) \,.
\end{equation}
For most pulsars, no glitches have been observed during O3,
and hence only the first term for the long-time spin-down evolution is relevant:
\begin{equation}
f_\mathrm{Taylor}(\tpsr)= \sum_{k=0}^{N} \frac{f^{(k)}(\tpsr-\Tref)^{k}}{k!},
\end{equation}
with a reference time $\Tref$
and up to $N$ frequency derivatives $f^{(k)}$.
We will also use $\fdot$, $\fddot$ and $\fdddot$
for the first three derivatives in this paper.
The three search methods in this paper have different maximum $N$ they can handle,
as discussed in Appendix~\ref{app:differing_ephemerides}.

For glitching pulsars, the additional term is
\begin{equation}
\begin{split}
f_\mathrm{glitch}(\tpsr)
 = \Theta(\tpsr-\Tgl)
   \left[
   \sum_{k=0}^{M} \frac{\Delta f^{(k)}_\glitch(\tpsr-\Tgl)^{k}}{k!}
   \right. \\
   \left. + \delta f_\mathrm{R} e^{-(\tpsr-\Tgl) / \tau_\mathrm{R}}\right]
\end{split}
\end{equation}
(or multiple such terms for pulsars with multiple glitches),
where $\Theta(x)$ is the Heaviside step function,
$\Tgl$ is the glitch time,
$\Delta f^{(k)}_\glitch$ with \mbox{$M \leq N$}
are the permanent jumps in the frequency and its derivatives,
$\delta f_\mathrm{R}$ is the part of the frequency jump that relaxes back,
and $\tau_\mathrm{R}$ is the relaxation time.
Relaxation is not necessarily observed for all glitches.
The glitch term is typically a small correction on top of $f_\mathrm{Taylor}(\tpsr)$.

Building on this frequency evolution model,
the full \gw signal received by a detector is
\begin{equation}
 \label{eq:hoft}
 h(\tdet) = F_+(\tdet;\alpha, \delta, \psi)h_+(\tdet) + F_\times(\tdet;\alpha, \delta, \psi)h_\times(\tdet),
\end{equation}
where \mbox{$F_{+,\times}(\tdet;\alpha, \delta, \psi)$} is the detector response
to $+$ and $\times$ polarized \gws
received at the detector at time $\tdet$,
coming from right ascension $\alpha$ and declination $\delta$
with polarization angle $\psi$~\citep{1998PhRvD..58f3001J}.
Timing corrections for translating from
$\tpsr$ to $\tdet$
are discussed below.

The amplitude of each \gw polarization can be written as
\begin{subequations}
 \label{eq:hplushcross}
 \begin{align}
  h_+(\tdet)      &= \frac{1}{2}h_0(1+\cos^2\iota)\cos\left[\phi_0 +\Phi(\tdet;\{f^{(k)}\},\alpha,\delta)\right], \label{eq:hplus} \\
  h_\times(\tdet) &= h_0\cos\iota\sin\left[\phi_0 +\Phi(\tdet;\{f^{(k)}\},\alpha,\delta)\right], \label{eq:hcross}
 \end{align}
\end{subequations}
where $\iota$ is the inclination angle of the pulsar, $\phi_0$ is the initial phase,
$h_0$ is an overall dimensionless strain amplitude as discussed below,
and \mbox{$\Phi(\tdet;\{f^{(k)}\},\alpha,\delta)$} is the integrated phase from the timing model of Eq.~\eqref{eq:frequency-model}.

To convert from $\tpsr$ at the source to $\tdet$ at the detector,
several timing corrections typically need to be addressed:
proper motion of the source, binary orbital motion, and the motion of the detectors around the solar system barycenter.
Proper motions between the pulsar and the solar system can usually be ignored~\citep{Covas:2020hcy} for these searches, and
in this analysis we do not cover sources in binary orbits as in, \textit{e.g.},
\citet{O3KnownPulsars,LIGOScientific:2020qhb,2021arXiv210909255T,AshokBeheshtipour2021}.
However, the sky position dependent correction between solar system barycenter and detector frame is a crucial part of the analysis methods described in the following, and is implemented in all cases~\citep{1998PhRvD..58f3001J}.

For \gws from a non-axisymmetric star, the amplitude $h_0$ can be written as
\begin{align}
    h_0 = \frac{4\pi^2G}{c^4}\frac{I_{zz}f^2\epsilon}{d},
\end{align}
where $d$ is the distance to the \ns,
$f$ is the \gw frequency,
and $I_{zz}$ is the principal moment of inertia.
The ellipticity, $\epsilon$, is given by
\begin{align}
    \epsilon = \frac{|I_{xx} - I_{yy}|}{I_{zz}}.
\end{align}

In summary, our \cw signal model depends on two sets of parameters,
the frequency-evolution parameters \mbox{$\Dop=\{f^{(k)},\alpha,\delta\}$},
and the amplitude parameters \mbox{$\Amp=\{h_0,\cos\iota,\psi,\phi_0\}$}.
The signal model for the long-duration transient search
is also based on the one discussed here,
but with an additional window function in time,
as will be discussed below in Sec.~\ref{sec:transient-method}.

For both \cw pipelines,
we search over a narrow range 
of \gw frequencies and first spin-down terms
centered on their respective central estimates from the source ephemeris:
\mbox{$\fgw = 2\fspin (1 \pm \kappa)$}
and \mbox{$\fdot = 2\fdotspin (1 \pm \kappa)$},
where we take \mbox{$\kappa=2\times10^{-3}$}.
This value is intended to allow for physical offsets between the frequency of \elmag-observed pulsar rotation and the \gw emission process, as discussed in the introduction and in more detail in \citet{2008ApJ...683L..45A,O2Narrowband}.

The frequency and spin-down changes due to glitches offer evidence of how large of a physical offset we might expect.
While glitches with
\mbox{$\Delta f^{(1)}_{\rm gl}/\dot{f}_{\rm spin} \gtrsim 2\times 10^{-3}$}
are observed occasionally, this value of $\kappa$ encompasses most glitches that have been observed \citep[see, \textit{e.g.},][]{Fuentes:2017bjx,Lower:2021rdo}
and increasing this value significantly to encompass all observed glitch sizes would not be computationally feasible.
The full widths we allow for both parameters,
\mbox{$\Delta \fgw=4\kappa\fspin$}
and \mbox{$\Delta\fdot=4\kappa\fdotspin$},
are shown in Table~\ref{tab:search_setup_parameter} for each target we consider.
Higher-order frequency derivatives are handled differently by each pipeline, which we discuss in the next two sub-sections and in Appendix~\ref{app:differing_ephemerides}.

\subsection{\texorpdfstring{$5n$}{5n}-vector narrowband pipeline}
\label{sec:fivevec-method}

The {\it $5n$-vector narrowband pipeline} uses the $5n$-vector method of \citet{2010CQGra..27s4016A,Astone:2012} to search for \cws in a narrow frequency and spin-down range around the source ephemerides.
This method searches for the characteristic modulations imprinted by the Earth's rotation on a GW signal,
splitting it into $5$ harmonics.
The search is applied for the $n$ detectors present in the network.
For more details on the implementation of the method we use here, see \citet{2017CQGra..34m5007M}. 

The starting point of the analysis is a collection of 1024\,s long \ffts
built as explained in the previous section.
For each target, we extract a frequency band large enough to contain the Doppler modulation of every template that we consider.
When higher-order frequency derivatives are used to fit the \elmag data,
the corresponding \gw spin-down terms (including $\fddot$) are fixed to twice the values provided in the source ephemerides.

For each target, we then correct for the Doppler modulation in the time domain using a non-uniform down-sampling to 1\,Hz.
Using this time-series, we then apply two matched filters (for the two \cw polarizations) and construct a detection statistic $S$ combining coherently the matched filter results from all detectors \citep{2017CQGra..34m5007M} for each template in the $\fgw$ and $\fdot$ space.
The detection statistic is defined as 
\begin{equation}
    S=|H_+|^2|A_+|^4+|H_\times|^2|A_\times|^4,
    \label{eq:s_statistic}
\end{equation}
where $A_{+/\times}$ are the Fourier components of $F_{+/\times}(t;\hat{n},\psi=0)$ and $H_{+/\times}$ are GW amplitude estimators built from the Fourier components of the data and $A_{+/\times}$.
The template bank bin size in frequency is given by the inverse of the time-series duration and the spin-down bin size is given by its inverse squared.

The final step is to select the local maximum of the detection statistic in every $10^{-4}$\,Hz band over all spin-down values considered.
Within this set of points in the parameter space, we select as outliers those with a $p$-value (defined as the tail probability of the detection statistic noise-only distribution) below a 1\% threshold, taking into account the number of templates applied in the \mbox{$\fgw$--$\fdot$} space.
The noise-only distribution of the detection statistic used to estimate the threshold for candidate selection is built using all the templates excluded from the analysis in each $10^{-4}$\,Hz band.
The detection statistic value corresponding to the threshold for candidate selection is extrapolated from the noise-only distribution using an exponential fit of the tail.

Finally, to compute the 95\% confidence level \uls $\hul$ on the $\cw$ amplitude in the case of no detection, we perform software injections with simulated \gw signals in each $10^{-4}$~Hz sub-band to estimate the value of $h_0$ at which we achieve 95\% detection efficiency at a false-alarm probability of 1\%. 

\subsection{Frequency-domain \texorpdfstring{\Fstat}{F-statistic} pipeline}
\label{sec:fstat-method}

We also use the frequency-domain \Fstat pipeline
to search for \cws in a narrow frequency and spin-down range around the source ephemerides.
The \Fstat is a matched filter which is maximized over the amplitude parameters $\Amp$ of the quasi-monochromatic signals~\citep{1998PhRvD..58f3001J,Cutler:2005hc};
see \citet{Tenorio:2021wmz} for a review covering its applications.

For each target we search over the range of \gw frequency $\fgw$ and spin-down $\fdot$,
with matched-filter templates chosen using the multi-detector \Fstat metric~\citep{Prix:2006wm,WettEtAl2008:SrGrvWvCssLI},
which we call through the \lalsuite~\citep{LALSuite} program \weave~\citep{WettEtAl2018:ImpSmSCnGrvWUOCnTB}.
If the second spin derivative $\fddotspin$ is non-zero,
we also search over the range \mbox{$\fddot = 2(\fddotspin \pm 3\sigma_{\fddotspin})$}
where $\sigma_{\fddotspin}$ is the $1\,\sigma$ uncertainty given in the source ephemeris.
Higher order frequency derivatives, up to $f^{(4)}$,
are fixed to the \elmag-measured values.
Because of constraints on the pipeline infrastructure,
if the source ephemeris for a target includes derivatives above $f^{(4)}$,
we refit the arrival times with fewer frequency derivatives.
This results in slightly different ephemerides being used for this search,
and for the $5n$-vector search.
We discuss targets for which this is the case in Appendix~\ref{app:differing_ephemerides}.

We place templates with a maximum matched-filter mismatch from a putative source of 0.02,
which leads to a very dense grid compared to the one used by the 5$n$-vector search.
This can be seen by comparing the final two (non-reference) columns of Table~\ref{tab:search_setup_parameter}.
The analysis uses a set of \SI{1800}{\second} Tukey-windowed \sfts over the full data span for all three detectors,
and produces a detection statistic, which we denote $2\F$, for each matched-filter template.
See Sec.~\ref{sec:data} for a full discussion of data processing and data quality cuts
made before generation of the \sfts.

Following \citet{Tenorio:2021wad},
we construct $10^{4}$ randomly-chosen batches of templates from our search results for each pulsar,
and fit a Gumbel distribution to the distribution of the maxima of these batches.
We can then propagate this Gumbel distribution based on the total number of templates 
to estimate the tail probability ($p$-value) of the largest template across the full search for that target.
We outline our implementation of this method more fully in Appendix~\ref{app:fstat_distromax_cw}.

If any templates have a $p$-value of less than 1\%
and corresponding large $2\F$ values,
we perform follow-up studies
of these templates.
A list of frequencies of known narrow spectral artifacts, $f_{\rm l}$, and their nominal widths, $w_{\rm l}$, are collated for all three detectors~\citep{linelist01, virgo3lines}.
If $|f_{\rm l} - f_{\rm large}|<w_{\rm l}$, where $f_{\rm large}$ is the frequency of the large-$2\F$ template,
then we veto that template.
We also look at the single-detector $2\F$ value calculated for each detector.
If the individual $2\F$ value for such a template in a single detector is larger than the $2\F$ value calculated using the whole network,
then we also veto that large-$2\F$ template~\citep{Keitel:2013wga,Leaci:2015iuc}.
Any remaining large-$2\F$ templates with \mbox{$p<1\%$} are then flagged for further follow-up,
and could be considered candidates for detection (subject to further studies).

In the absence of a detection, we set \uls at 95\% confidence, $\hul$,
for each target using the software-injection scheme set out in, \textit{e.g.}, \citet{2007PhRvD..76h2001A}.

\subsection{Post-glitch transient search}
\label{sec:transient-method}

The search for long-duration transient CW-like signals
from glitching pulsars is motivated by the idea
that some of the excess energy liberated in a glitch
could correspond to transient changes in the quadrupole moment
of the star and be radiated away in \gws~\citep{vanEysden:2008pd,Bennett:2010tm,Prix:2011qv,Melatos:2015oca,Singh:2016ilt,Yim:2020trr}
during the post-glitch relaxation phase.

To search for such signals,
we use the transient \Fstat method introduced by \citet{Prix:2011qv}
and previously applied to LIGO O2 data in searches for \gws
from glitches in the Crab and Vela pulsars~\citep{Keitel:2019zhb}.
The idea is to search for long-duration transients
that are ``\cw-like'' in the sense of
following the standard quasi-monochromatic \cw signal model from Sec.~\ref{sec:signal}
with only an additional window function $\win(\tdet;\tzero,\tau)$ in time applied:
\begin{equation}
 \label{eq:htCW}
 h(\tdet;\Dop,\Amp,\TP) = \win(\tdet;\tzero,\tau)\,h(\tdet;\Dop,\Amp) \,,
\end{equation}
where the transient parameters $\TP$ consist of
the window shape,
signal start time $\tzero$,
and a duration parameter $\tau$.
As in~\citet{Keitel:2019zhb}, we limit ourselves here to the simplest case
of rectangular windows, meaning the signal is exactly a standard \cw
that starts at time $\tzero$ and is cut off at time $\tzero+\tau$,
with no additional amplitude evolution.
Some form of amplitude decay would be expected for a realistic
signal linked to post-glitch relaxation~\citep{Yim:2020trr}.
As demonstrated in \citet{Prix:2011qv} and \citet{Keitel:2019zhb},
the \snr loss from using rectangular windows in a search
for exponentially decaying signals is mild,
while using exponentially windowed search templates
would mean a much higher computational cost~\citep{Keitel:2018pxz}.

\begin{figure*}
    \centering
    \includegraphics[width=0.9\textwidth]{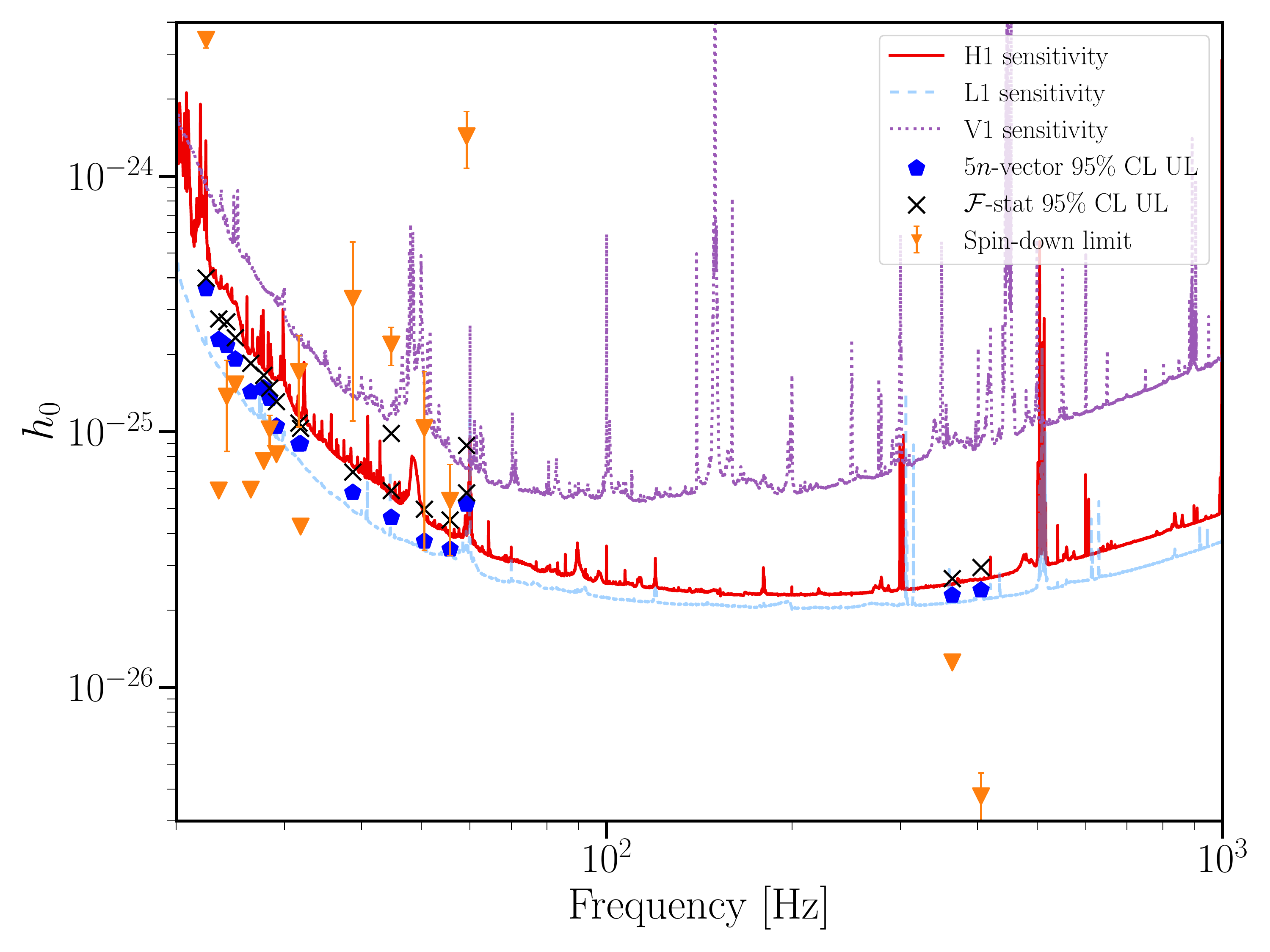}
    \caption{
    \label{fig:ULs}
    The red solid, blue dashed, and purple dotted curves show the expected sensitivities for H1, L1, and V1 respectively,
    using Eq.~\ref{eq:sens-estimate}.
    The blue pentagons indicate the median 95\% CL \uls from the 5$n$-vector search across all $10^{-4}$~Hz sub-bands for each source.
    The black crosses indicate 95\% CL \uls from the \Fstat search, which are set across the full search range for each target.
    The orange triangles indicate the spin-down limit, $\hsd$, with error bars that reflect uncertainty in the distance to each source.
    In a few cases the error bars are smaller than the size of the markers.
    We discuss and compare these limits in more detail in Sec.~\ref{sec:conclusion}.
    We do not show uncertainties on \uls here, although we discuss uncertainties due to the \ul-setting method as well as calibration uncertainties in Sec.~\ref{sec:conclusion}.
    The data for this figure are available online~\citep{data_release}.
    }
\end{figure*}

The search uses the same underlying \Fstat library functions in \lalsuite~\citep{LALSuite}
as the \cw search described in Sec.~\ref{sec:fstat-method},
analyzing \sft data
with the \cfs program
which was used before, \textit{e.g.}, in searches for \cws from supernova remnants~\citep{LIGOScientific:2014ymk,2019ApJ...875..122A,Lindblom:2020rug}.
For each target, we perform a search at fixed sky location
over a grid in $f^{(k)}$ parameters,
with the number of spin-down terms depending on the pulsar ephemerides,
going up to at most a $\fdddot$ term.
At each grid point $\Dop$,
the standard multi-detector \Fstat is calculated over the maximum duration of interest.
Intermediate per-\sft quantities from this calculation,
the so-called \Fstat atoms,
are kept.
As described in \citet{Prix:2011qv}, partial sums of these atoms are evaluated in a loop over a range of $\{\tzero,\tau\}$.
The resulting matrix of $\Fmn=\F(t_{0m},\tau_n)$ statistics
gives the likelihoods of transient \cw-like signals in each time range.
We marginalize $\Fmn$ over uniform priors on $\tzero$ and $\tau$,
obtaining the Bayes factor $\Btrans$ for transient \cw-like signals against Gaussian noise
as our detection statistic.
As demonstrated by \citet{Prix:2011qv},
$\Btrans$ improves detection efficiency
compared to taking the maximum of $\Fmn$,
and as demonstrated by \citet{Tenorio:2021wad},
it also produces cleaner background distributions on real data.
The computational cost of these searches scales linearly with the number of $\Dop$ templates
and with the product of the numbers of $\tzero$ and $\tau$ values,
and is dominated by the partial summing steps over the latter two parameters.
See \citet{Prix:2011qv} and \citet{Keitel:2018pxz} for details.

As this is only the second search thus far for long-transient \gws of this type \citep[after][]{Keitel:2019zhb}, we describe its practical setup in more detail in Sec.~\ref{sec:transient-setup}.

\section{CW search results}
\label{sec:results}

\begin{deluxetable*}{lccccccccccc}
\tablecaption{
 \ul results on \cw strain from the 5$n$-vector (5v) and \Fstat ($\F$) pipelines.
 \label{tab:upper_limits_results_tab}
}
\tablehead{\colhead{Name} & \colhead{$f$} & \colhead{$h_0^{95}$ (5v)} & \colhead{$h_0^{95}$ ($\F$)} & \colhead{$h_{\rm sd}$} & \colhead{$h_0^{95} / h_{\rm sd}$} & \colhead{$h_0^{95} / h_{\rm sd}$} & \colhead{$\epsilon$ (5v)} & \colhead{$\epsilon$ ($\F$)} & \colhead{$\epsilon_{\rm sd}$} & \colhead{$d$} & \colhead{$\sigma_d$} \\
\colhead{} & \colhead{Hz} & \colhead{$\times10^{-26}$} & \colhead{$\times10^{-26}$} & \colhead{$\times10^{-26}$}& \colhead{(5v)}& \colhead{ ($\F$)}& \colhead{$\times10^{-5}$}& \colhead{$\times10^{-5}$}& \colhead{$\times10^{-5}$}& \colhead{kpc} &\colhead{kpc}}
\startdata
J0534+2200 bg   & 59.24  & ...  & 8.79  & 143.0  & ...   & 0.061  & ...    & 4.6    & 76.0    & 2.00$\;^{a}$ & 0.5 \\
J0534+2200 ag   & 59.24  & 5.2  & 6.09  & 143.0  & 0.036 & 0.043  & 2.7    & 3.2    & 76.0    & 2.00$\;^{a}$ & 0.5 \\
J0711--6830      & 364.23 & 2.29 & 2.69  & 1.25   & 1.8   & 2.2    & 0.0017 & 0.002  & 0.00095 & 0.11 & 0.041 \\
J0835--4510      & 22.37  & 36.2 & 39.5  & 341.0  & 0.11  & 0.12   & 19.0   & 21.0   & 180.0   & 0.28$\;^{b}$ & 0.02 \\
J1101--6101      & 31.85  & 8.96 & 10.5  & 4.25   & 2.1   & 2.5    & 59.0   & 68.0   & 28.0    & 7.00 & 2.7 \\
J1105--6107      & 31.64  & 8.94 & 10.1  & 17.1   & 0.52  & 0.59   & 20.3   & 23.0   & 38.0    & 2.36 & 0.92 \\
J1809--1917      & 24.17  & 21.7 & 26.1  & 13.7   & 1.6   & 1.9    & 110.0  & 140.0  & 73.0    & 3.27 & 1.3 \\
J1813--1749 full & 44.70  & 4.61 & 5.85  & 21.9   & 0.21  & 0.27   & 10.0   & 13.0   & 64.0    & 6.20$\;^{c}$ & 2.4 \\
J1813--1749 bg   & 44.70  & ...  & 9.61  & 21.9   & ...   & 0.44   & ...    & 21.0   & 64.0    & 6.20$\;^{c}$ & 2.4 \\
J1828--1101      & 27.75  & 14.8 & 16.3  & 7.67   & 1.9   & 2.1    & 87.0   & 96.0   & 45.0    & 4.77 & 1.9 \\
J1833--0827      & 23.45  & 22.9 & 29.3  & 5.88   & 3.9   & 5.0    & 180.0  & 230.0  & 46.0    & 4.50 & 1.8 \\
J1838--0655      & 28.36  & 13.2 & 15.3  & 10.2   & 1.3   & 1.5    & 100.0  & 120.0  & 79.0    & 6.60$\;^{d}$ & 0.9 \\
J1856+0245      & 19.35  & 20.5 & ...   & 11.2   & 1.8   & ...    & 200.0  & ...    & 110.0   & 6.32 & 2.46 \\
J1913+1011      & 55.69  & 3.47 & 4.37  & 5.36   & 0.65  & 0.81   & 4.9    & 6.1    & 7.5     & 4.61 & 1.8 \\
J1925+1720      & 26.43  & 14.3 & 17.6  & 5.93   & 2.4   & 3.0    & 98.0   & 120.0  & 41.0    & 5.06 & 2.0 \\
J1928+1746      & 29.10  & 10.5 & 13.5  & 8.15   & 1.3   & 1.7    & 51.0   & 66.0   & 39.0    & 4.34 & 1.7 \\
J1935+2025      & 24.95  & 19.2 & 23.7  & 15.3   & 1.3   & 1.5    & 130.0  & 170.0  & 110.0   & 4.60 & 1.8 \\
J1952+3252      & 50.59  & 3.73 & 4.93  & 10.3   & 0.36  & 0.48   & 4.1    & 5.5    & 11.0    & 3.00$\;^{e}$ & 2.0 \\
J2124--3358      & 405.59 & 2.4  & 2.9   & 0.374  & 6.4   & 7.7    & 0.0057 & 0.0068 & 0.00095 & 0.44$\;^{f}$ & 0.05 \\
J2229+6114      & 38.71  & 5.78 & 7.23  & 33.1   & 0.17  & 0.22   & 11.0   & 14.0   & 63.0    & 3.00$\;^{g}$ & 2.0
\enddata
\tablecomments{
We show the \uls on strain amplitude at 95\% confidence for both pipelines, $h_0^{95}$, the spin-down limit for each target $h_{\rm sd}$ along with the implied limit on ellipticity $\epsilon$ for each pipeline, and the limit on ellipticity implied by the spin-down limit $\epsilon_{\rm sd}$.
Finally, we also show the ratio of the 95\% confidence \ul and the spin-down limit.
We surpass the spin-down limit for \numnarrowbandexceedspindwon targets:  \psr{0534+2200}, \psr{0835-4510}, \psr{1105--6107}, \psr{1813--1749}, \psr{1913+1011}, \psr{1952+3252}, and \psr{2229+6114}.
We do not show uncertainty on \uls here, although we discuss uncertainty due to the \ul-setting method as well as calibration uncertainty in Sec.~\ref{sec:conclusion}.
Distance estimates $d$ and their uncertainties $\sigma_d$ are from dispersion measures \citep[with a fiducial uncertainty of 40\% from][]{2017ApJ...835...29Y} unless noted with one of the following: (a)~\citet{Kaplan:2008qm} (b)~\citet{2003ApJ...596.1137D} (c)~\citet{CamiloRansom2021} (d)~\citet{GotthelfHalpern2008} (e)~\citet{2012ApJ...755...39V} (f)~\citet{ReardonShannon2021} (g)~\citet{HalpernCamilo2001}.
This table and a machine-readable file are available online~\citep{data_release}.
}
\end{deluxetable*}

As described below, we find no significant candidates, although both the \Fstat and $5n$-vector \cw searches find outliers requiring further follow up after the vetoes in Sec.~\ref{sec:model-methods} were applied.
Hence, we set observational \uls on the \gw strain from each target,
constraining \gw emission below the spin-down limit on \numnarrowbandexceedspindwon of the \numnarrowbandtargets target pulsars.
All \ul results are shown in Fig.~\ref{fig:ULs},
and listed in Table~\ref{tab:upper_limits_results_tab}.
We give details of the results from both pipelines,
including outliers that were followed up,
in the rest of this section.

When discussing \uls in the following section,
it should be noted that physically meaningful constraints on the GW energy emission
are set when the \uls are lower than the spin-down limit
(i.e., when the limit has been \emph{surpassed}).
Ellipticity constraints for pulsars whose spin-down limit is not surpassed would imply a $|\fdotspin|$ higher than the one observed. 

Finally, distance estimates used for the spin-down limits are given in Table~\ref{tab:upper_limits_results_tab}.
These are either from the ATNF pulsar catalogue
(based on dispersion measures),
or from the literature.
For the pulsar \psr{1813--1749}, different models of the electron density in the Galaxy yield different distance estimates~\citep{CamiloRansom2021}.
We have used the more optimistic estimate $d = 6.2\pm2.4\,{\rm kpc}$,
although a different model gives a more pessimistic estimate of $d = \SI{12}{kpc}$.

\subsection{\texorpdfstring{$5n$}{5n}-vector narrowband pipeline}
\label{sec:fivevec-results}

\begin{figure}
    \centering
    \includegraphics[width=0.45\textwidth]{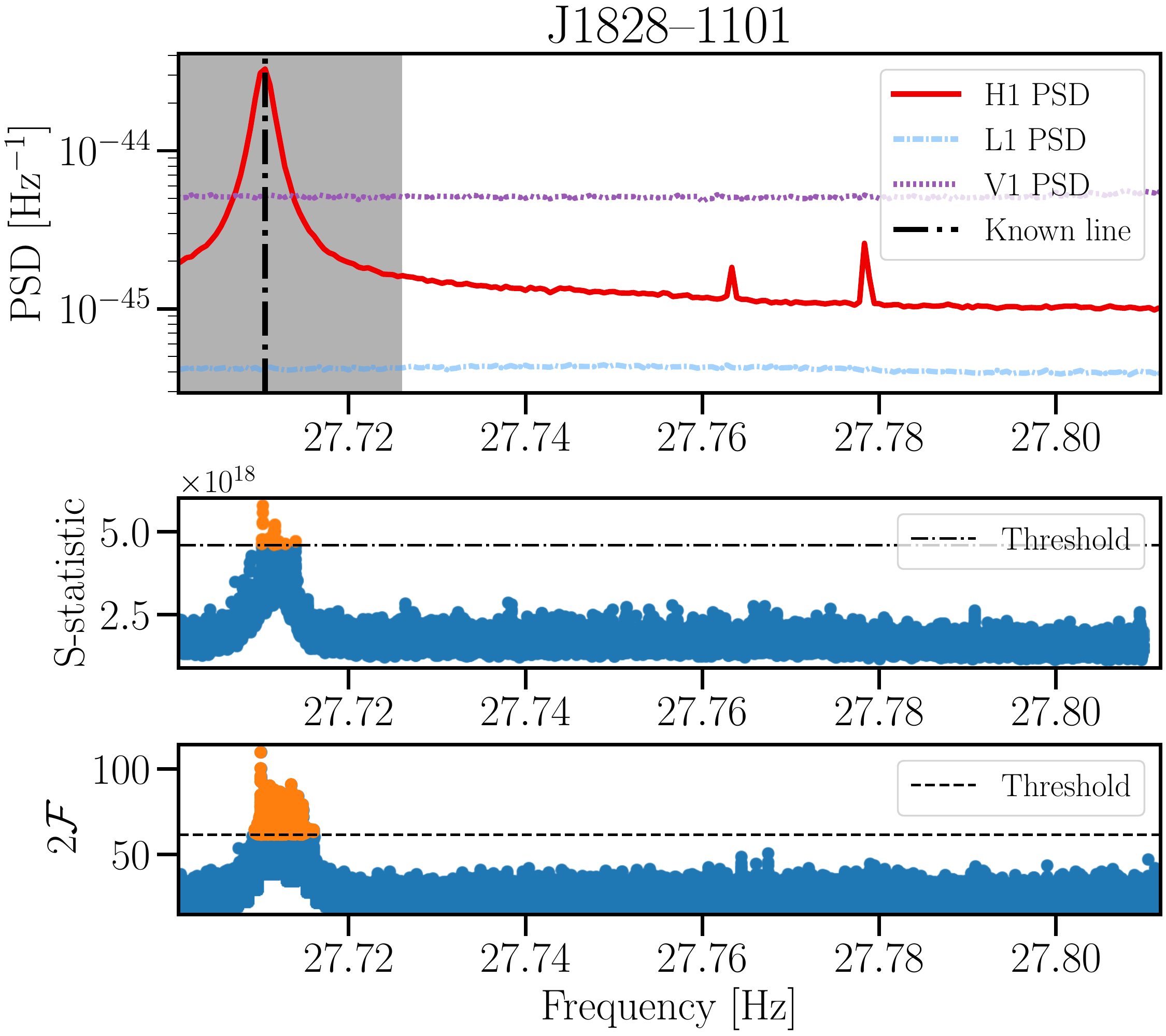}
    \caption{
    \label{fig:J1828m1101_S_candidates}
    \textit{Top panel}: Median \psd across the run in each frequency bin for Hanford (red, solid), Livingston (blue, dashed), and Virgo (purple, dotted) for the frequency band searched for \psr{1828--1101}.
    There is an obvious disturbance caused by a resonance in one of the suspended optics at H1, which is marked with the vertical dashed black line.
    Its estimated width is given by the shaded region.
    The line is clearly associated with large values of the detection statistics in the two other panels.
    \textit{Middle panel}: $S$-statistic obtained for \psr{1828--1101} in the narrow frequency range explored and marginalized over the spin-down values. The horizontal dashed line indicates the threshold corresponding to a $1\%$ false-alarm rate. The orange markers indicate the outliers found.
    \textit{Bottom panel}: $2\F$ obtained for \psr{1828--1101} in the narrow frequency range explored.
    The horizontal dashed line indicates the threshold corresponding to a $1\%$ false-alarm rate.
    The orange markers indicate the outliers found and vetoed using the known lines veto.
    }
\end{figure}

\begin{figure}
    \centering
    \includegraphics[width=0.45\textwidth]{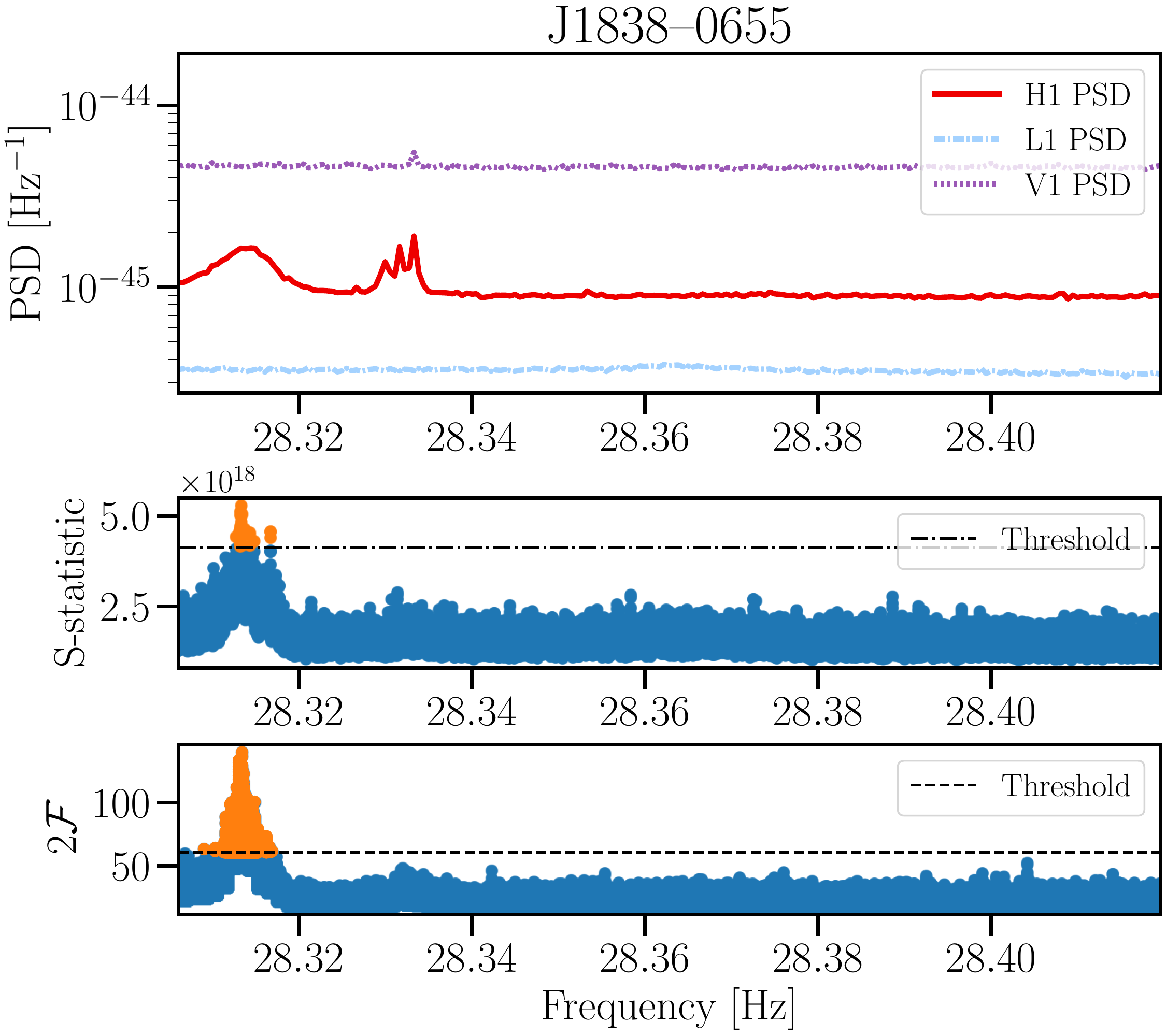}
    \caption{
    \label{fig:J1838m0655_S_candidates}
    \textit{Top panel}: Median \psd across the run in each frequency bin for Hanford (red, solid), Livingston (blue, dashed), and Virgo (purple, dotted) for the frequency band searched for \psr{1838--0655}.
    There is an obvious disturbance (of unknown origin) at H1 which is associated with large values of the detection statistics in the two other panels.
    \textit{Middle panel}: $S$-statistic obtained for \psr{1838--1101} in the narrow frequency range explored and marginalized over the spin-down values.
    The horizontal dashed line indicates the threshold corresponding to a $1\%$ false-alarm rate.
    The orange markers indicate the outliers found.
    \textit{Bottom panel}: $2\F$ obtained for \psr{1838--1101} in the narrow frequency range explored.
    The horizontal dashed line indicates the threshold corresponding to a $1\%$ false alarm rate.
    The orange markers indicate the outliers found.
    Follow-ups of these outliers are discussed in Appendix~\ref{app:final_follow_up}.
    }
\end{figure}

The 5$n$-vector pipeline found outliers only for two targets: \psr{1828--1101} and \psr{1838--0655}.
Figures~\ref{fig:J1828m1101_S_candidates} and \ref{fig:J1838m0655_S_candidates} show the distribution of outliers in the searched frequency band (marginalized over the spin-down plane), along with the \psd for each of the three detectors in that band.
For \psr{1828--1101} we find 10 outliers around 27.7116 Hz with $p$-values between $10^{-5}-10^{-2}$,
and for \psr{1838--0655} we find 13 outliers around 28.3134 Hz with $p$-values between $10^{-3}-10^{-2}$.
The nominal $p$-values we report here assume underlying Gaussian noise, an assumption that can break down in the presence of instrumental artifacts.

\begin{figure}
    \centering
    \includegraphics[width=0.45\textwidth]{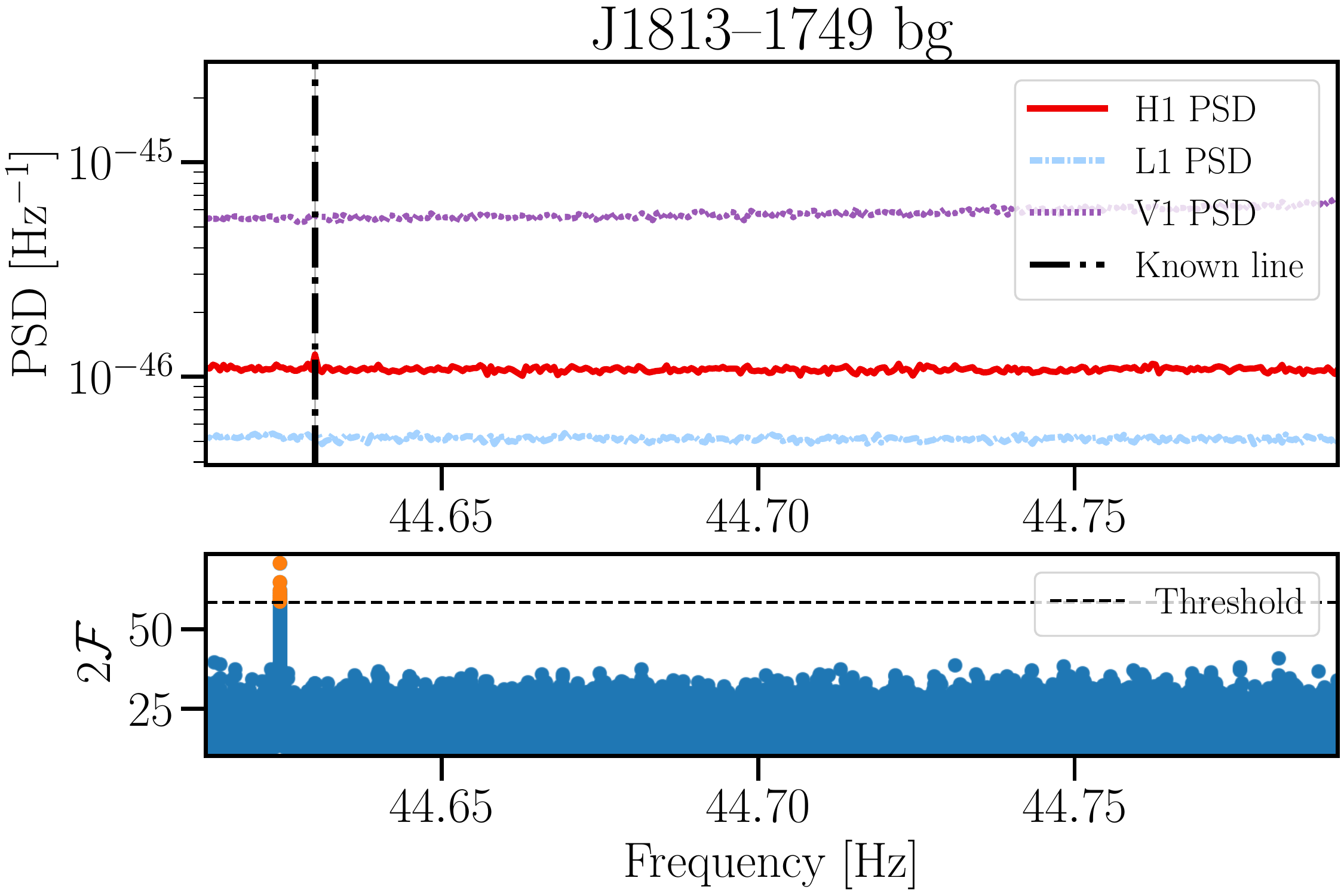}
    \caption{
    \label{fig:fstat_J1813-1749_results}
    \textit{Top panel}: Median \psd across the run in each frequency bin for Hanford (red), Livingston (blue), and Virgo (purple)
    for the frequency band searched for \psr{1813--1749}.
    There is a small bump near a known line in H1 (marked with the vertical dashed line).
    However, this is distinct from the spike in $2\F$ values shown in the bottom panel.
    \textit{Bottom panel}: $2\F$ obtained for \psr{1813--1749} bg in the narrow frequency range explored.
    The horizontal dashed line indicates the threshold corresponding to a $1\%$ false-alarm rate.
    The orange markers indicate the outliers found.
    We were unable to veto this explicitly due to the known line, but after follow-up detailed in Appendix~\ref{app:final_follow_up} it is clear that these outliers are caused by an artifact that affects the first two weeks of \ohthree.
    }
\end{figure}

These outliers are all due to noise disturbances identified in H1 data. For \psr{1828--1101}, the PSD for each of the three detectors around the outliers are shown in the top panel of Fig.~\ref{fig:J1828m1101_S_candidates}, while the $S$-statistic, defined in Eq.~\ref{eq:s_statistic}, is shown in the middle panel. The outliers are produced by a known noise line at \SI{27.71}{\hertz} with width of 0.02 Hz caused by a resonance in one of the suspended optics at H1.    
For \psr{1838--0655} the outliers are due to an identified broad noise line of unknown origin~\citep{linelist01}.
The top panel of Fig. \ref{fig:J1838m0655_S_candidates} shows the \psd for each of the three detectors around these outliers.
As can be seen, various broad noise disturbances are present in H1 data in the frequency band of \psr{1838--0655}.
Even if the origin of these noise lines is unknown, we can confidently exclude them as astrophysical \cw signals since, if they were real, they would have been present also in the L1 data which is more sensitive. 
In Sec.~\ref{sec:fstat-results} we also perform a dedicated follow-up for the candidates due to these noise disturbances showing that they are only visible in H1 data.

The median \uls across the $10^{-4}~\rm{Hz}$ sub-bands, $\hul$, for each pulsar are shown in column 3 of Table~\ref{tab:upper_limits_results_tab} and plotted with blue pentagons in Fig.~\ref{fig:ULs}.
The ratio of the \ul with $\hsd$ is shown in column 6 of Tab.~\ref{tab:upper_limits_results_tab}, and the limits on the ellipticity of the pulsar are shown in column 8. We discuss and compare these limits to those of the \Fstat pipeline, to previous searches for the same targets, and to the spin-down limits in Sections~\ref{sec:pipeline-comparison} and \ref{sec:conclusion}.

\subsection{Frequency-domain \texorpdfstring{\Fstat}{F-statistic} pipeline}
\label{sec:fstat-results}
The \Fstat pipeline identified outliers with $p$-value $<0.01$ that could not be vetoed with the methods described in Sec.~\ref{sec:fstat-method} for two targets, \psr{1813--1749} bg and \psr{1838--0655}.
As described in Sec.~\ref{sec:fstat-method}, a first round of vetoes was made where we rejected outliers that were close to known narrow spectral artifacts. This resulted in vetoing outliers in \psr{1828--1101}, which are shown in the bottom panel of Fig.~\ref{fig:J1828m1101_S_candidates} for completeness and comparison with the $5n$-vector search.
We also vetoed outliers with a $2\F$ value that was larger running on data from a single detector than when running on the full network.
In Figs.~\ref{fig:J1838m0655_S_candidates} and \ref{fig:fstat_J1813-1749_results} we show the \psd in the top panel in the search band for the remaining outliers for \psr{1838--0655} and \psr{1813--1749} bg respectively, while the search pipeline statistic, the \Fstat discussed in Sec.~\ref{sec:fstat-method}, is shown in the bottom panel(s).

The outliers for \psr{1813--1749} bg are associated with an artifact that contaminates the first two weeks of \ohthree. While nearby, it is outside the nominal width of the line specified in~\citet{linelist01}, which can be seen in Figure~\ref{fig:fstat_J1813-1749_results}, and is therefore unlikely to be explicitly caused by that line.
However, in Appendix~\ref{app:final_follow_up}, we perform a follow-up study that shows that the outliers in this band increase in significance when running on data only from the first two weeks of the run.
This behavior is inconsistent with a \cw signal, and so we veto these outliers.

The remaining outliers from \psr{1838--0655}, shown in Fig.~\ref{fig:J1838m0655_S_candidates}, are in the same frequency range as those for the 5$n$-vector pipeline.
These outliers are all near a clear disturbance in H1, and the $2\F$ value for all of them is larger when running only on H1 data than when running only on L1 data.
However, L1 is significantly more sensitive than H1 in the frequency bands of interest.
In Appendix~\ref{app:final_follow_up} we detail a follow-up study on the remaining outliers.
We show, based on a realistic set of injections,
that for a true signal it is very unlikely to have a larger $2\F$ value when running only on H1 data than when running only on L1 data,
due to the difference in the sensitivity for the two detectors.

One pulsar, \psr{1856+0245}, was in a frequency band that had significant instrumental disturbances, and so we could not reliably estimate the background distribution, and do not report \uls (or search results) for it.
We discuss this case in Appendix~\ref{app:fstat_distromax_cw}.

Given that there are no remaining candidates that exceed our follow-up threshold, we set \uls at 95\% confidence on the strain amplitude, $h_0$, of a potential \cw signal produced by our targets.
The \ul, calculated across the full parameter space, $\hul$, for each pulsar is shown in column 4 of Table~\ref{tab:upper_limits_results_tab} and plotted with black crosses in Fig.~\ref{fig:ULs}.
The ratio $\hul/\hsd$ is shown in column 7 of Tab.~\ref{tab:upper_limits_results_tab}, and the limit on the ellipticity of each pulsar is shown in column 9.
We discuss these limits in detail, and in the context of indirect limits and the 5$n$-vector results, in Section~\ref{sec:conclusion}.

\subsection{Comparison between pipelines}
\label{sec:pipeline-comparison}

The \uls from the \Fstat pipeline are, on average, 19\% higher than those set by the 5$n$-vector pipeline.
They are not directly comparable, however, as the 5$n$-vector pipeline results are the median across $10^{-4}~\rm{Hz}$ sub-bands, while the \Fstat limits are set across the whole parameter space for each pulsar.
A more equitable comparison is to compare the largest \ul across all sub-bands for the 5$n$-vector, which is on average 9\% lower than the \uls set by the \Fstat pipeline. 

A direct comparison would be very involved due to the differing pre-treatment of the data and methods used by the two pipelines.
However, there are a few clear methodological differences, which could explain the difference in \uls, which we highlight below.

The most obvious difference is in the effective ``trials factors'' used in generating the thresholds for follow-up, which are then used for the threshold of ``detection'' when performing the software injections used to calculate \uls.
Looking at a larger number of templates reduces the effective mismatch between templates and a signal, but also increases the likelihood of finding a larger detection statistic due to a noise fluctuation.
The \Fstat search uses significantly more templates than the 5$n$-vector pipeline in nearly all cases. 
Fitting the relationship between the ratio of templates used in the two pipelines to the ratio of their \uls, we find that for an equal number of templates, the \Fstat search would have limits $\sim 5\%$ higher than the maximum \ul compared to the 5$n$-vector search.
This is consistent with the scaling found in~\citet{AstoneColla2014}. 
Inevitably, these differences are a reflection of the full range of choices made for the two pipelines, which also include how thresholds for follow-up are set
(\textit{e.g.}, the method outlined in Appendix~\ref{app:fstat_distromax_cw} versus extrapolating the tail of the background distribution),
the cleaning procedures used in preparing the data, as well as the densities of the templates used to perform the searches.

\section{Post-glitch transient search}
\label{sec:transient}

\subsection{Search setup}
\label{sec:transient-setup}

For all \numglitchtargets targets in the post-glitch transient searches,
we use \sfts of duration 1800\,s
(as discussed in Sec.~\ref{sec:data})
from both LIGO detectors and Virgo,
except for \ohnineoheight and \eighteentwentysix.
The frequency bands searched for these two targets are affected by broad-band instrumental disturbances in Virgo.
For this analysis,
we additionally clean
narrow instrumental lines with known instrumental origin~\citep{linelist01,virgo3lines}
before the searches
by replacing affected \sft bins with Gaussian noise matching the \psd in the surrounding range.

\begin{figure}[t]
 \includegraphics[width=\columnwidth]{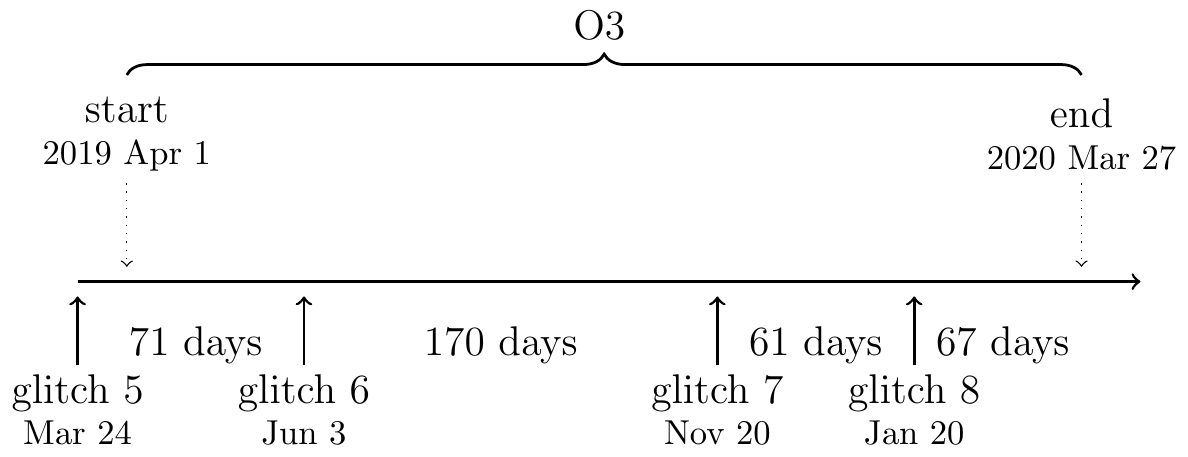}
 \caption{
  \label{fig:transient-J0537}
  Relative timing of
  the O3 observing run,
  the glitches that \ac{nicer} observed from \ohfivethirtyseven~\citep{Ho:2020vxt,LIGOScientific:2020lkw},
  and the four post-glitch long-duration transient searches we perform for this target.
 }
\end{figure}

Observed parameters and search setups for the \numglitchtargetswithoutbigglitcher pulsars that have a single known glitch during our time of interest
are summarized in Table~\ref{tab:transientPSRs}.
For the ``Big Glitcher'' \ohfivethirtyseven~\citep{Middleditch:2006ky,Ho:2020vxt},
which was previously considered in searches for persistent \cws in the inter-glitch periods
by \citet{Fesik:2020tvn} and \citet{LIGOScientific:2020lkw,LIGOScientific:2021yby},
we perform four separate searches following four observed glitches as illustrated in Fig.~\ref{fig:transient-J0537}.
Corresponding parameters are listed in Table~\ref{tab:transientJ0537}.

\begin{deluxetable*}{lccccc}
\tablecaption{
 \label{tab:transientPSRs}
 Parameters and transient search setups
 for \numglitchtargetswithoutbigglitcher of the glitching pulsars.
}
\tablehead{ & \colhead{J0534+2200} & \colhead{J0908--4913} & \colhead{J1105--6107} & \colhead{J1813--1749} & \colhead{J1826--1334}}
\startdata
$d$ [pc] & $2000 \pm 500$ \tablenotemark{a} & $1000 \pm 390$ & $2360 \pm 920.4$ & $6200 \pm 2400$ \tablenotemark{b} & $3606 \pm 1406$ \\
R.A. & $05^{\textrm{h}}34^{\textrm{m}}31.97^{\textrm{s}}$ & $09^{\textrm{h}}08^{\textrm{m}}35.47^{\textrm{s}}$ & $11^{\textrm{h}}05^{\textrm{m}}25.71^{\textrm{s}}$ & $18^{\textrm{h}}13^{\textrm{m}}35.11^{\textrm{s}}$ & $18^{\textrm{h}}26^{\textrm{m}}13.16^{\textrm{s}}$ \\
DEC. & $22^\circ00'52.07''$ & $-49^\circ13'05.00''$ & $-61^\circ07'55.63''$ & $-17^\circ49'57.57''$ & $-13^\circ34'45.98''$ \\
$\Tgl$ [s] & 1247930924 & 1254619602 & 1238824009 & 1248825618 & 1264809618 \\
$\Delta \Tgl$ [$\days$] & 0.003 & 4.488 & 1.997 & 1 & 21 \\
$\glstepf/f$ & $\hphantom{-}1.24\times10^{-8}$ & $\hphantom{-}2.21\times10^{-8}$ & $\hphantom{-}1.17\times10^{-6}$ & $\hphantom{-}1.34\times10^{-7}$ & $\hphantom{-}2.48\times10^{-6}$ \\
$\glstepfdot/\fdot$ & $\hphantom{-}1.52\times10^{-4}$ & $\hphantom{-}6.73\times10^{-4}$ & $\hphantom{-}3.53\times10^{-3}$ & $\dots$ & $\hphantom{-}6.82\times10^{-3}$ \\
$\glstepfddot/\fddot$ & $\dots$ & $\dots$ & $\dots$ & $\dots$ & $\dots$ \\
$\Tref$ [s] & $\hphantom{-}1247845509$ & $\hphantom{-}1256655802$ & $\hphantom{-}1238651637$ & $\hphantom{-}1248739243$ & $\hphantom{-}1262996751$ \\
$f$ [Hz] & $\hphantom{-} 59.204190$ & $\hphantom{-} 18.722385$ & $\hphantom{-} 31.628009$ & $\hphantom{-} 44.679639$ & $\hphantom{-} 19.691348$ \\
$\Delta f$ [Hz] & $\hphantom{-}0.059234$ & $\hphantom{-}0.018732$ & $\hphantom{-}0.031644$ & $\hphantom{-}0.044702$ & $\hphantom{-}0.019701$ \\
$\fdot$ [Hz\,s$^{-1}$] & $-7.3707\times10^{-10}$ & $-2.6526\times10^{-12}$ & $-7.9786\times10^{-12}$ & $-1.2863\times10^{-10}$ & $-1.4650\times10^{-11}$ \\
$\Delta\fdot$ [Hz\,s$^{-1}$] & $\hphantom{-}7.3670\times10^{-13}$ & $\hphantom{-}2.6513\times10^{-15}$ & $\hphantom{-}3.5865\times10^{-14}$ & $\hphantom{-}1.3200\times10^{-13}$ & $\hphantom{-}1.1358\times10^{-13}$ \\
$\fddot$ [Hz\,s$^{-2}$] & $\hphantom{-}2.3631\times10^{-20}$ & $\hphantom{-}\dots$ & $\hphantom{-}6.2723\times10^{-21}$ & $\hphantom{-}\dots$ & $\hphantom{-}8.9844\times10^{-22}$ \\
$\Delta\fddot$ [Hz\,s$^{-2}$] & $\hphantom{-}2.3643\times10^{-23}$ & $\hphantom{-}\dots$ & $\hphantom{-}1.1809\times10^{-21}$ & $\hphantom{-}\dots$ & $\hphantom{-}7.6842\times10^{-24}$ \\
$\fdddot$ [Hz\,s$^{-3}$] & $\hphantom{-}\dots$ & $\hphantom{-}\dots$ & $-5.1143\times10^{-28}$ & $\hphantom{-}\dots$ & $\hphantom{-}4.2734\times10^{-29}$ \\
$\Delta\fdddot$ [Hz\,s$^{-3}$] & $\hphantom{-}\dots$ & $\hphantom{-}\dots$ & $\hphantom{-}7.9821\times10^{-29}$ & $\hphantom{-}\dots$ & $\hphantom{-}2.3250\times10^{-30}$ \\
$\Ntemp$ & 3917139 & 507911 & 61497 & 7120630 & 22753 \\
\enddata

\tablecomments{
 This table contains information about
 the \numglitchtargetswithoutbigglitcher pulsars with a single post-glitch transient analysis.
 See Table~\ref{tab:transientJ0537} for the special case of \ohfivethirtyseven.
 Distances are taken from the ATNF catalog
 unless indicated otherwise with a footnote,
 and all other parameters are derived from the ephemerides
 as discussed in the text.
 $\glstepf$, $\glstepfdot$ and $\glstepfddot$ are the glitch step sizes.
 For each search,
 the reference time $\Tref$ is chosen to match the earliest \sft data timestamp
 in $\Tgl\pm\min(\Delta\Tgl,1\,\mathrm{d})$,
 and the \gw frequency and spin-down parameters $f$, $\fdot$ etc.
 are extrapolated to this time.
 $\Delta f$, $\Delta\fdot$ etc. refer to their corresponding search band widths.
 $\Ntemp$ is the total number of frequency-evolution parameters covered by each search.
}
\tablenotetext{a}{\cite{Trimble:1973pasp,Kaplan:2008qm}}
\tablenotetext{b}{\cite{CamiloRansom2021}}

\end{deluxetable*}

\begin{deluxetable*}{lcccccc}
\tablecaption{
 \label{tab:transientJ0537}
 Parameters of the four searches targeting glitches of \ohfivethirtyseven.
}
\tablehead{ & \colhead{glitch5} & \colhead{glitch6} & \colhead{glitch7} & \colhead{glitch8}}
\startdata
$\Tgl$ [s] & 1237420818 & 1243555218 & 1258243218 & 1263513618 \\
$\Delta \Tgl$ [$\days$] & 5 & 8 & 3 & 5 \\
$\glstepf/f$ & $\hphantom{-}1.49\times10^{-7}$ & $\hphantom{-}4.36\times10^{-7}$ & $\hphantom{-}1.22\times10^{-7}$ & $\hphantom{-}3.88\times10^{-7}$ \\
$\glstepfdot/\fdot$ & $\hphantom{-}4.77\times10^{-4}$ & $\hphantom{-}4.34\times10^{-4}$ & $\hphantom{-}1.11\times10^{-3}$ & $\hphantom{-}1.22\times10^{-3}$ \\
$\glstepfddot/\fddot$ & $\hphantom{-}2.00\times10^{-1}$ & $-1.34\times10^{-1}$ & $\hphantom{-}4.88$ & $-4.32$ \\
$\Tref$ [s] & $\hphantom{-}1238166018$ & $\hphantom{-}1242864932$ & $\hphantom{-}1257985627$ & $\hphantom{-}1263081781$ \\
$f$ [Hz] & $\hphantom{-}123.767904$ & $\hphantom{-}123.766064$ & $\hphantom{-}123.760042$ & $\hphantom{-}123.758055$ \\
$\Delta f$ [Hz] & $\hphantom{-}0.123830$ & $\hphantom{-}0.123828$ & $\hphantom{-}0.123822$ & $\hphantom{-}0.123820$ \\
$\fdot$ [Hz\,s$^{-1}$] & $-3.9990\times10^{-10}$ & $-3.9980\times10^{-10}$ & $-4.0034\times10^{-10}$ & $-4.0017\times10^{-10}$ \\
$\Delta\fdot$ [Hz\,s$^{-1}$] & $\hphantom{-}3.9970\times10^{-13}$ & $\hphantom{-}3.9960\times10^{-13}$ & $\hphantom{-}7.3810\times10^{-13}$ & $\hphantom{-}7.3493\times10^{-13}$ \\
$\fddot$ [Hz\,s$^{-2}$] & $\hphantom{-}1.9990\times10^{-20}$ & $\hphantom{-}1.4420\times10^{-20}$ & $\hphantom{-}9.7927\times10^{-20}$ & $-9.4572\times10^{-20}$ \\
$\Delta\fddot$ [Hz\,s$^{-2}$] & $\hphantom{-}2.0000\times10^{-23}$ & $\hphantom{-}3.7517\times10^{-21}$ & $\hphantom{-}1.4036\times10^{-19}$ & $\hphantom{-}1.1735\times10^{-19}$ \\
$\fdddot$ [Hz\,s$^{-3}$] & $\hphantom{-}\dots$ & $\hphantom{-}\dots$ & $\hphantom{-}\dots$ & $\hphantom{-}\dots$ \\
$\Delta\fdddot$ [Hz\,s$^{-3}$] & $\hphantom{-}\dots$ & $\hphantom{-}\dots$ & $\hphantom{-}\dots$ & $\hphantom{-}\dots$ \\
$\Ntemp$ & 593911 & 11474334 & 15837589 & 21543992 \\
\enddata

\tablecomments{
 Here, $d=(49.59\pm0.55)$\,kpc~\citep{2019Natur.567..200P},
 the right ascension is $05^{\mathrm{h}}37^{\mathrm{m}}47.42^{\mathrm{s}}$
 and the declination is $69^{\circ}10'19.88''$
 for all four searches,
 while the frequency and spin-down parameters are extrapolated
 to the $\Tref$ corresponding to each search.
 See Table~\ref{tab:transientPSRs} for details on the listed parameters.
}
\end{deluxetable*}

We derive the ranges of
frequency evolution parameters $\Dop$ and transient parameters $\TP$
to be covered in each search
from the ephemerides for that pulsar,
the uncertainty $\delta\Tgl$ in the glitch time $\Tgl$,
and the availability of \gw data relative to $\Tgl$.
We set a maximum signal duration of $\tau_{\max}=120\,\days$
to be searched for each glitch,
guided both by
observed glitch recovery times being typically on shorter timescales \citep{Lyne:2000sta,2011MNRAS.414.1679E,2013MNRAS.429..688Y,Yim:2020trr,Lower:2021rdo}
and by computational cost constraints~\citep{Keitel:2018pxz}.
Signal start times $\tzero$ cover $\Tgl\pm\max(\delta\Tgl,1\,\mathrm{d})$
and the reference time $\Tref$ for the frequency and spin-down parameter grids
(again assuming \gw emission near twice the rotation period)
is set to the earliest \sft timestamp in this range.
On each side of the nominal $f^{(k)}$ values extrapolated to $\Tref$,
the grids cover the maximum of
(i) 0.001 times the nominal value
(this is one quarter of the width used by the \cw searches as described in Sec.~\ref{sec:signal}),
(ii) the 1$\sigma$ fit uncertainties from the ephemerides propagated to $\Tref$,
and (iii) the glitch step size in that parameter.
We place the grids using the metric from \citet{Prix:2006wm}
and the template placement algorithm from \citet{Wette:2014tca},
as implemented in \cfs and \code{lalpulsar}~\citep{LALSuite},
with a mismatch parameter of 0.2.
The algorithm can add some additional grid points outside the nominal ranges
to reduce mismatch near the edges.
But for targets with $\fddot$ and higher terms in their timing solution,
we strictly limit template placement in all spin-down terms to the nominal range,
to limit computational cost.
Our resolution in $\TP$ is $\dtzero=\dtau=2\,\Tsft=3600\,$s.
Due to implementation details of the \Fstat~\citep{Prix:2015cfs},
the minimum duration is
$\tau_{\min}=2\,\Tsft$.

In some special cases we modified the default setup to match the availability of \gw data,
explaining some apparent anomalies in the configurations listed in Tables \ref{tab:transientPSRs} and \ref{tab:transientJ0537}.
The glitch from \ohnineoheight happened in October of 2019,
during the maintenance break between O3a and O3b,
a month for which no \gw data is available.
We still search for the standard set of $\tau$ up to 120\,days after this glitch,
but fix $\tzero$ to the first available SFT timestamp after the break --
any templates with a different $\tzero$ in the regular band around the glitch time would yield identical detection statistics --
and are insensitive to shorter-duration signals in this case.
For \eighteentwentysix,
O3 ended 84 days after the nominal glitch time,
so we shorten the analysis accordingly.

For \ohfivethirtyseven,
as illustrated in Fig.~\ref{fig:transient-J0537}
and following the numbering from \citet{Ho:2020vxt},
we perform searches covering the durations
from each of glitches 5--7
until the next glitch,
and for glitch 8 we search until the end of O3.
Glitch 5 happened in late March of 2019, shortly before the start of O3,
and hence, as for \ohnineoheight,
$\tzero$ is fixed to the data start time.

\subsection{Results}
\label{sec:transient-results}

To determine whether or not there are interesting outliers in the results of each search,
for most targets we again follow the procedure described in~\citet{Tenorio:2021wad}
to empirically estimate the distribution of the expected highest outlier.
For two targets,
\elevenohfive and \eighteentwentysix,
the number of templates is too low for this method to deliver robust results.
Instead, for these targets we use spatial off-sourcing~\citep{Isi:2020uxj} to estimate the background distribution and set a threshold.
The specific implementations for both cases are explained in Appendix~\ref{app:fstat_distromax_transient}.

\begin{figure*}[thp]
 \includegraphics[width=\textwidth]{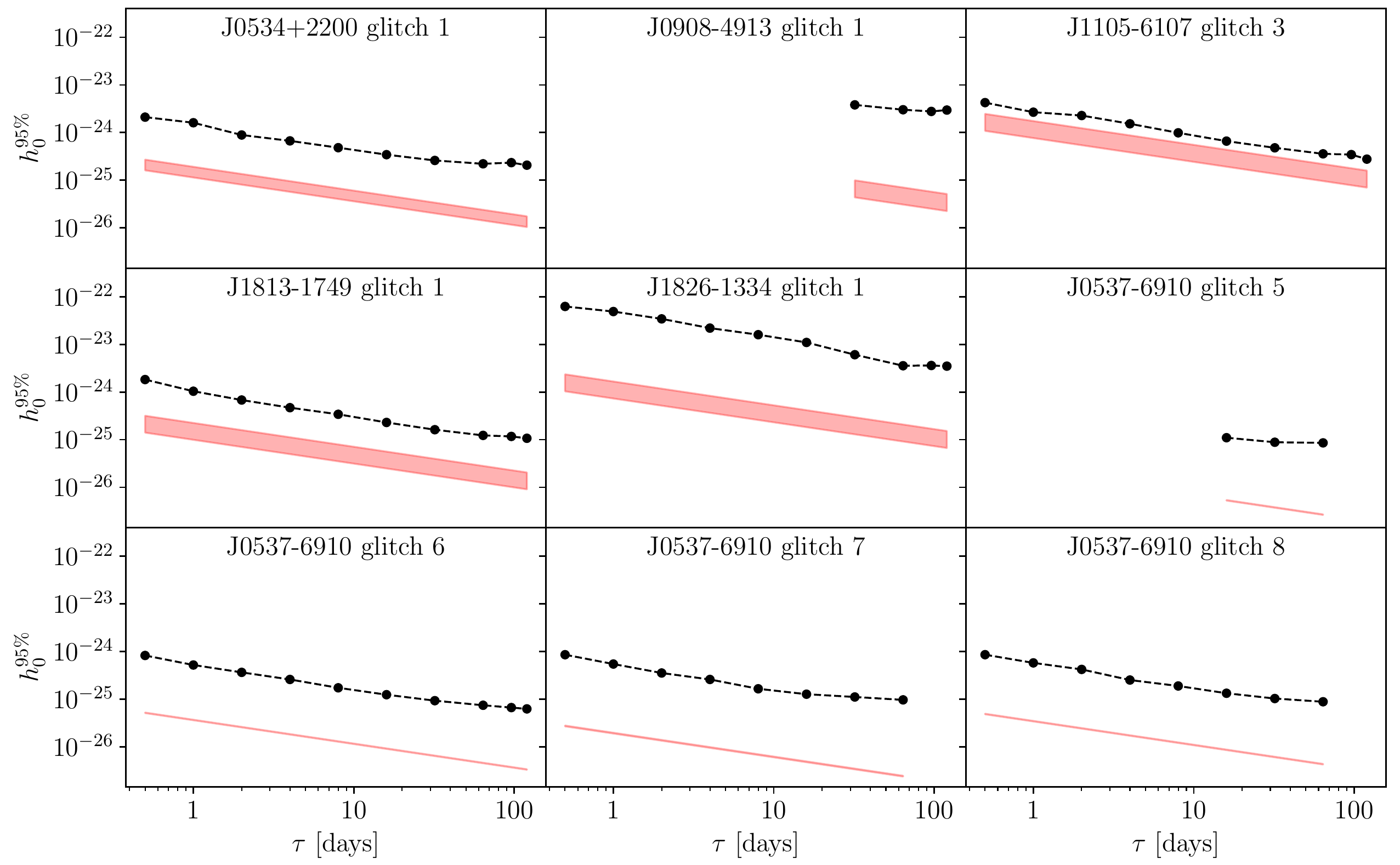}
 \caption{
  \label{fig:transient-ULs}
  Strain \uls $\hul$ obtained for the \numglitchtargets glitching pulsars
  targeted with the long-duration transient search,
  as a function of signal duration $\tau$
  (dashed curves with points
  representing the injection set).
  The four searches following glitches in \ohfivethirtyseven are shown in separate panels
  for readability.
  Uncertainties from finite injection sample size are at the 3--5\% level,
  comparable to or lower than
  amplitude uncertainty from detector calibration
  as discussed in Sec.~\ref{sec:conclusion}.
  The indirect glitch excess energy \uls
  from Eq.~\eqref{eq:glitchenergyUL},
  originally derived in \citet{Prix:2011qv}
  and based on the estimated size of each glitch
  and distance to the pulsar as listed in
  Tables \ref{tab:transientPSRs} and \ref{tab:transientJ0537},
  are shown for comparison
  (red shaded bands).
  The data for this figure are available online~\citep{data_release}.
 }
\end{figure*}

For \numglitchsearcheswithnooutliers of the \numglitchsearches searches,
there are no candidates above threshold.
The background distribution from the search after the eighth glitch of \ohfivethirtyseven
is less clean and requires an additional notching step,
as discussed in Appendix~\ref{app:fstat_distromax_transient}.
We then find two marginal outliers
at about 2 and 1 standard deviations respectively from the mean of the estimated extreme-value distribution (4\% and 14\% tail probability).
The best-fit signal durations are about 60 and 45 days, respectively.
Appendix~\ref{app:J0537-outliers}
lists the full parameters of these outliers
and describes additional analyses to follow them up.

In summary, we cannot associate either outlier with any known spectral artifact
or single-detector feature in the data.
They pass several consistency tests and vetoes,
and behave as weak \cw-like candidates
are expected to in a fully phase-coherent targeted follow-up~\citep{Dupuis:2005,2017arXiv170508978P}
over the corresponding subsets of O3 data.
Nevertheless, we do not consider these as promising detection candidates
for two main reasons:
first, they would imply a signal with far higher amplitude
than allowed by the estimated glitch energy for this target
(see discussion of \uls below).
Second, their significance is low,
compared also with the 1\% tail probability thresholds used by the two \cw pipelines in this paper.
This is reinforced by follow-up $5n$-vector fixed-duration analyses,
which recover local $S$-statistic peaks at the parameters of each outlier,
but not as the loudest candidates in the band.
Still, we conservatively use a higher threshold
(from including the outliers in the background estimate with no notching)
in the \ul procedure for \ohfivethirtyseven glitch 8.

Having identified no convincing detection candidates above threshold
for any of the targets,
we set \uls on the \gw strain after each glitch.
As in the search, we assume CW-like signals that are constant over the signal duration $\tau$,
and we evaluate the \uls as a function of $\tau$.
To do so, we simulate signals following Eq.~\eqref{eq:htCW}
with $\Dop$ parameters and $\tzero$ drawn uniformly from within the search ranges,
several discrete steps in $\tau$ and $h_0$,
and the remaining amplitude parameters randomized over their natural ranges.
We add these signals to the original \sfts without the spectral cleaning (as described in Sec.~\ref{sec:transient-setup}),
redo the cleaning step to ensure
any signals falsely dismissed due to cleaning are properly accounted for,
and perform small searches over a portion of the original $\Dop$ grid
(0.1\,mHz in $\fgw$ and the closest spin-down points)
and the full $\TP$ range.
The $\hul$ \uls are obtained from a sigmoid fit to the
recovery fraction $\pdet$ above the $\tCWdetstat$ threshold from the main search,
as a function of injected $h_0$ at fixed $\tau$.
Results for all targets are shown in Fig.~\ref{fig:transient-ULs}.

We compare these empirical \uls
to an indirect constraint derived in~\citet{Prix:2011qv}
according to the two-fluid model of \ns glitches~\citep{Lyne:2000sta}.
In this model, the excess superfluid energy liberated in a glitch
gives a \ul to the total \gw energy,
which corresponds to a $\tau$-dependent strain \ul
\begin{equation}
 \label{eq:glitchenergyUL}
 h_0 \leq \frac{1}{d} \sqrt{\frac{5G}{2c^3} \frac{\Izz}{\tau} \frac{\glstepf}{f}} \,.
\end{equation}
Qualitatively similar \uls also hold for other glitch models
such as crust-cracking starquakes~\citep{Middleditch:2006ky};
see again~\citet{Prix:2011qv} for details.
For showing this indirect \ul in Fig.~\ref{fig:transient-ULs},
we have assumed a fiducial moment of inertia
\mbox{$\Izz=10^{38}$\,kg\,m$^2$}
and distances to each target as listed in Tables \ref{tab:transientPSRs} and \ref{tab:transientJ0537}.

As can be see in Fig.~\ref{fig:transient-ULs},
our observational \uls do not reach below the indirect energy constraint
for any of the targets.
While the sensitivity of \ohthree was improved over O2,
none of the \ohthree targets had such a favorable combination of large frequency step size and small distance as the one from the Vela pulsar during O2~\citep{Palfreyman:2016gli,Sarkissian:2017awh,Palfreyman:2018gli}.
Our closest result to the benchmark is for \elevenohfive,
which had the second largest glitch of the selected targets
in terms of $\glstepf/f$,
and is at a closer distance and more favorable frequency than
the only other target with a larger glitch (\eighteentwentysix).
For this glitch, the closest the $\tau$-dependent \uls get to the indirect energy constraint is within a factor of 1.6.
As discussed, these empirical \uls are valid specifically for \gw signals
following the simple \cw-like model of Eq.~\eqref{eq:htCW},
whereas many other models for \gw emission following pulsar glitches are conceivable.

\section{Conclusion}
\label{sec:conclusion}

We have used data from the LIGO--Virgo observing run \ohthree
to search for \cw signals from \numnarrowbandtargets known pulsars,
allowing for a mismatch between the frequencies of \elmag and \gw emission,
and for long-duration \cw-like transients from \numglitchtargets glitching pulsars.
All statistically significant outliers were associated with instrumental disturbances and vetoed, and so 
we set \uls on the \gw strain amplitude from each target.

The narrowband \cw results obtained here
complement the more constraining limits obtained from perfectly targeted searches on the same \ohthree data~\citep{O3KnownPulsars},
where the \gw frequency was assumed to be at exactly once or twice
the \elmag-measured rotation frequency.
With respect to the narrowband search done with the first half of O3 \citep{LIGOScientific:2020gml}, we have improved our \uls by a factor of $\sim 35\%$ for \psr{0534+2200} and $\sim 10\%$ for \psr{0711--6830} and \psr{0835--4510}, using a parameter space $\sim 6.5, 1.2$ and $7.1$ larger for these three pulsars.
For the remaining pulsars which were also searched for using the same method in O2, the limits presented here are a factor of 2--3 lower than those set in~\citet{O2Narrowband}.
We surpass the spin-down limits for \numnarrowbandexceedspindwon pulsars,
including \psr{1105--6107} and \psr{1913+1011},
for the first time using the narrowband method.

The best upper limit set across the whole frequency band is $h_{0}\lesssim 2.3\times 10^{-26}$ for \psr{0711--6830} at roughly 364\,Hz using the $5n$-vector search.
We also set the best limit on pulsar ellipticity with \psr{0711--6830} at $\epsilon \lesssim 1.7\times10^{-8}$.
However, neither of these limits surpass the indirect spin-down limits for this pulsar, which are a factor of $\sim 2.2$ lower than the upper limits for the search. For pulsars with $\hul<\hsd$, the best $\hul$ is $h_0\lesssim 3.5\times10^{-26}$ for \psr{1913+1011} and best ellipticity limit is $\epsilon \lesssim 2.7\times10^{-5}$ for \crab.

Our long-duration transient search results for signals following \numglitchsearches glitches in \numglitchtargets pulsars
significantly extend the target set for such searches
over the two glitches from two pulsars previously covered
with O2 data by~\citet{Keitel:2019zhb},
demonstrating the flexibility of the search method.
We report two marginal outliers from the search for the last glitch of \ohfivethirtyseven during O3,
but do not consider them as significant detection candidates.
We have set duration-dependent strain \uls for each post-glitch search
and compared the results with the indirect glitch energy \ul benchmark established in~\citet{Prix:2011qv},
but not reached this benchmark
for any of these targets,
with \elevenohfive coming closest to it.

Although we do not explicitly show systematic error or statistical uncertainty on upper limits presented in Figs.~\ref{fig:ULs} and \ref{fig:transient-ULs} or Tab.~\ref{tab:upper_limits_results_tab}, their impact on these results is discussed here.
Statistical uncertainty due to finite sampling when calculating limits is
$\sim 3\%$ for the \Fstat CW search
and 3--5\% for the post-glitch transient search.
For the 5$n$-vector search, uncertainty due to which part of the frequency band is sampled
(i.e., typical distance from median \ul across all 0.1 mHz sub-bands)
is $\sim 2.4$\%. 
Meanwhile, systematic error in the calibration of the O3 strain data is quantified in~\citet{Sun:2020wke,Sun:2021qcg} for LIGO and~\citet{VIRGO:2021kfv} for Virgo.
The error depends upon the time at which data was collected and upon the frequency band of the measurement. 
The maximum estimated systematic error in absolute magnitude of the strain across all frequencies and time is 7\% for the LIGO detectors in O3a, 11\% for the LIGO detectors in O3b, and 5\% for Virgo,
though typically the error is smaller than these levels.
Additionally, calibration systematics are not correlated between detectors,
and so it is likely that the absolute error in our combined measurements would be less than these values.

We expect that with future upgrades of the LIGO--Virgo--KAGRA network~\citep{2020LRR....23....3A}
we can further improve \gw results on the known pulsar population,
and constrain emission scenarios and nuclear physics
both for quiescent-state pulsars and those perturbed by glitches.

\section*{Acknowledgements}
This material is based upon work supported by NSF’s LIGO Laboratory which is a major facility
fully funded by the National Science Foundation.
The authors also gratefully acknowledge the support of
the Science and Technology Facilities Council (STFC) of the
United Kingdom, the Max-Planck-Society (MPS), and the State of
Niedersachsen/Germany for support of the construction of Advanced LIGO 
and construction and operation of the GEO\,600 detector. 
Additional support for Advanced LIGO was provided by the Australian Research Council.
The authors gratefully acknowledge the Italian Istituto Nazionale di Fisica Nucleare (INFN),  
the French Centre National de la Recherche Scientifique (CNRS) and
the Netherlands Organization for Scientific Research (NWO), 
for the construction and operation of the Virgo detector
and the creation and support  of the EGO consortium. 
The authors also gratefully acknowledge research support from these agencies as well as by 
the Council of Scientific and Industrial Research of India, 
the Department of Science and Technology, India,
the Science \& Engineering Research Board (SERB), India,
the Ministry of Human Resource Development, India,
the Spanish Agencia Estatal de Investigaci\'on (AEI),
the Spanish Ministerio de Ciencia e Innovaci\'on and Ministerio de Universidades,
the Conselleria de Fons Europeus, Universitat i Cultura and the Direcci\'o General de Pol\'{\i}tica Universitaria i Recerca del Govern de les Illes Balears,
the Conselleria d'Innovaci\'o, Universitats, Ci\`encia i Societat Digital de la Generalitat Valenciana and
the CERCA Programme Generalitat de Catalunya, Spain,
the National Science Centre of Poland and the European Union – European Regional Development Fund; Foundation for Polish Science (FNP),
the Swiss National Science Foundation (SNSF),
the Russian Foundation for Basic Research, 
the Russian Science Foundation,
the European Commission,
the European Social Funds (ESF),
the European Regional Development Funds (ERDF),
the Royal Society, 
the Scottish Funding Council, 
the Scottish Universities Physics Alliance, 
the Hungarian Scientific Research Fund (OTKA),
the French Lyon Institute of Origins (LIO),
the Belgian Fonds de la Recherche Scientifique (FRS-FNRS), 
Actions de Recherche Concertées (ARC) and
Fonds Wetenschappelijk Onderzoek – Vlaanderen (FWO), Belgium,
the Paris \^{I}le-de-France Region, 
the National Research, Development and Innovation Office Hungary (NKFIH), 
the National Research Foundation of Korea,
the Natural Science and Engineering Research Council Canada,
Canadian Foundation for Innovation (CFI),
the Brazilian Ministry of Science, Technology, and Innovations,
the International Center for Theoretical Physics South American Institute for Fundamental Research (ICTP-SAIFR), 
the Research Grants Council of Hong Kong,
the National Natural Science Foundation of China (NSFC),
the Leverhulme Trust, 
the Research Corporation, 
the Ministry of Science and Technology (MOST), Taiwan,
the United States Department of Energy,
and
the Kavli Foundation.
The authors gratefully acknowledge the support of the NSF, STFC, INFN and CNRS for provision of computational resources.

This work was supported by MEXT, JSPS Leading-edge Research Infrastructure Program, JSPS Grant-in-Aid for Specially Promoted Research 26000005, JSPS Grant-in-Aid for Scientific Research on Innovative Areas 2905: JP17H06358, JP17H06361 and JP17H06364, JSPS Core-to-Core Program A. Advanced Research Networks, JSPS Grant-in-Aid for Scientific Research (S) 17H06133 and 20H05639 , JSPS Grant-in-Aid for Transformative Research Areas (A) 20A203: JP20H05854, the joint research program of the Institute for Cosmic Ray Research, University of Tokyo, National Research Foundation (NRF), Computing Infrastructure Project of KISTI-GSDC, Korea Astronomy and Space Science Institute (KASI), and Ministry of Science and ICT (MSIT) in Korea, Academia Sinica (AS), AS Grid Center (ASGC) and the Ministry of Science and Technology (MoST) in Taiwan under grants including AS-CDA-105-M06, Advanced Technology Center (ATC) of NAOJ, and Mechanical Engineering Center of KEK.


{\it We would like to thank all of the essential workers who put their health at risk during the COVID-19 pandemic, without whom we would not have been able to complete this work.}

We acknowledge that CHIME is located on the traditional, ancestral, and unceded territory of the Syilx/Okanagan people. We are grateful to the staff of the Dominion Radio Astrophysical Observatory, which is operated by the National Research Council of Canada.  CHIME is funded by a grant from the Canada Foundation for Innovation (CFI) 2012 Leading Edge Fund (Project 31170) and by contributions from the provinces of British Columbia, Qu\'{e}bec and Ontario. The CHIME/FRB Project, which enabled development in common with the CHIME/Pulsar instrument, is funded by a grant from the CFI 2015 Innovation Fund (Project 33213) and by contributions from the provinces of British Columbia and Qu\'{e}bec, and by the Dunlap Institute for Astronomy and Astrophysics at the University of Toronto. Additional support was provided by the Canadian Institute for Advanced Research (CIFAR), McGill University and the McGill Space Institute thanks to the Trottier Family Foundation, and the University of British Columbia. The CHIME/Pulsar instrument hardware was funded by NSERC RTI-1 grant EQPEQ 458893-2014. This research was enabled in part by support provided by WestGrid (\url{www.westgrid.ca}) and Compute Canada (\url{www.computecanada.ca}). 
We acknowledge support from the Natural Sciences and Engineering Research Council of Canada (NSERC) funding reference \#CITA 490888-16, the Canadian Institute for Advanced Research, and the UBC Four Year Fellowship (6456).

We acknowledge support from EPSRC/STFC fellowship (EP/T017325/1), ANID/FONDECYT grants 1171421 and 1211964, and NASA grants 80NSSC19K1444 and 80NSSC21K0091. This work is supported by NASA through the NICER mission and the Astrophysics Explorers Program and uses data and software provided by the High Energy Astrophysics Science Archive Research Center (HEASARC), which is a service of the Astrophysics Science Division at NASA/GSFC and High Energy Astrophysics Division of the Smithsonian Astrophysical Observatory.

\software{
\code{Astropy}~\citep{astropy:2013,astropy:2018},
\code{corner.py}~\citep{corner},
\code{distromax}~\citep{distromax,Tenorio:2021wad},
\code{GWpy}~\citep{gwpy},
\code{LALSuite}~\citep{LALSuite},
\code{matplotlib}~\citep{matplotlib},
\code{numpy}~\citep{harris2020array},
\code{OctApps}~\citep{2018JOSS....3..707W},
\code{pandas}~\citep{pandas,mckinney-proc-scipy-2010},
\code{ptemcee}~\citep{Foreman-Mackey:2012any,Vousden:2015pte},
\pyfstat~\citep{Ashton:2018ure,pyfstat,Keitel:2021xeq},
\code{scipy}~\citep{scipy},
\code{SWIGLAL}~\citep{Wette:2020air},
\code{Tempo}~\citep{NiceDemorest2015},
\code{Tempo2}~\citep{Edwards:2006zg}.
}

\bibliographystyle{aasjournal}

\appendix

\section{Differences in pulsar ephemerides used for different pipelines}
\label{app:differing_ephemerides}

Due to variations in code infrastructure for the \Fstat and 5$n$-vector pipelines, there were some cases where we used ephemerides with different parameterizations to search for the same source.
In general, the \weave code used for the \Fstat CW search can search up to the fourth derivative in frequency,
and the automatic template placement will work best if the reference time at which the frequency and spin-downs are defined is during or close to the time at which we are conducting the search.
Similarly, the transient search (also based on the \Fstat, but using the \cfs code) can search up to $\fdddot$.
On the other hand, the 5$n$-vector pipeline has no such limitations on the number of frequency derivatives.

In several cases, which we document below, the \elmag ephemerides used for the 5$n$-vector search included higher-order frequency derivatives to improve the phenomenological fit and reduce the effects of timing noise.
In those cases, we refit the ephemerides using at most a fourth frequency derivative for the \Fstat pipelines and we then used the simple heuristic from Eq. 11 of~\citet{PrixItoh2005} to verify that the effect of the increased timing residuals on our matched-filter analysis was less than the maximum mismatch used to place templates for the search.
The affected pulsars were:
\begin{itemize}
    \item \psr{0534+2200}: The Crab pulsar was originally fit with twelve frequency derivatives, with an average timing residual of \texttt{TRES} $=135.0~\rm{\mu s}$, and this ephemeris was used for the 5$n$-vector search.
    For both \Fstat-based pipelines we search with up to two frequency derivatives with an average timing residual of \texttt{TRES} $=482.382~\rm{\mu s}$.
    \item \psr{0835--4510}:
    The Vela pulsar was originally fit with seven frequency derivatives, with an average timing residual of \texttt{TRES} $=137.1~\rm{\mu s}$, and this ephemeris was used for the 5$n$-vector search.
    The \Fstat CW pipeline ran with a search up to four frequency derivatives with an average timing residual of  \texttt{TRES} $=133.2~\rm{\mu s}$.
    \item \psr{1809--1917}:
    This pulsar was originally fit with five frequency derivatives, with an average timing residual of \texttt{TRES} $=1766.4~\rm{\mu s}$, and this ephemeris was used for the 5$n$-vector search.
    The \Fstat pipeline ran with a search up to four frequency derivatives with an average timing residual of \texttt{TRES} $=489.4~\rm{\mu s}$.
    In this case, it is likely that the timing residuals are smaller because the reference time used for the fits was moved to the middle of the observation time.
\end{itemize}

\section{settings for background estimation in \texorpdfstring{$\F$}{F}-statistic pipelines}
\label{app:fstat_distromax}

\subsection{CW search}
\label{app:fstat_distromax_cw}

In this appendix, we outline the details of the method used for estimating the distribution of the expected largest outlier for the frequency-domain \Fstat \cw search. We then set a threshold corresponding to a $p$-value of 1\% under the assumption that no signal is present in the data.
We use the \dmax method introduced by~\citet{Tenorio:2021wad}, with some slight modifications due to the specific situation at hand.

We want to estimate the distribution of the maximum value of $2\F$ for the full search for a single pulsar, under the assumption that no signal is present.
To be conservative, we will \textit{exclude} bands corresponding to known disturbances when estimating this distribution, and as such we will have outliers that exceed our threshold to follow up that are associated with the disturbances.
However, this means that we will not set our threshold accounting for the disturbances, which would generally push them upwards and potentially miss a signal.
The procedure is as follows:
\begin{enumerate}
    \item Run the search and
    save the templates with the top $2\times10^{7}$ values of $2\F$.
    \item Excise templates within the established width of known lines in~\citet{linelist01,virgo3lines}.
    \item Remove templates that are within $4\times10^{-5}f$ of unidentified lines, and in the case where the line is specified as ``broad'' in~\citet{linelist01} we remove templates within $4\times 10^{-4}f$.
    \item Perform a single iteration of the notching procedure from \citet{Tenorio:2021wad}, which removes templates with a frequency within $5\times 10^{-5}$\,Hz of any template exceeding an internal threshold. 
    \item Split the remaining $2\F$ values into $10^{4}$ random batches of size $\sim 2\times 10^3$.
    Find the maximum $2\F$ in each of those batches, and fit a Gumbel distribution to that set of maxima.
    \item Propagate that Gumbel distribution based on the number of batches~\citep{Tenorio:2021wad}, and the \textit{total} number of templates
    (i.e., the number of templates used before we perform notching, since this is the number of templates used to perform the search)
    to obtain a distribution for the maximum of the full search for that individual pulsar.
\end{enumerate}

What is left is a distribution on the maximum value of $2\F$ under the assumption that no signal is present, and we can then integrate this distribution to find the value of $2\F$ that corresponds to a $p$-value of 1\%.
Any of the $2\times10^{7}$ original templates with $p<$1\% are then subject to first the vetoing procedure described in Sec.~\ref{sec:fstat-method}, and if they pass this procedure, then they are flagged for more extensive follow-up.

With the exception of one pulsar, \psr{1856+0245}, the notching procedure above removes less than 40\% of the frequency band.
In the case of \psr{1856+0245}, the known and unknown lines notching procedure removes the full frequency band, and so we do not search for CWs from this pulsar with the $\F$-statistic pipeline.

\subsection{Transient search}
\label{app:fstat_distromax_transient}

The basic approach for most targets in this analysis is the same as for the \cw search,
following the \dmax method introduced by \citet{Tenorio:2021wad}:
we fit a Gumbel distribution to the measured maxima
of our detection statistic $\tCWdetstat$
over random subsets (``batchmaxes'').
We then propagate the parameters of this distribution considering the full number of templates.
However, we then set the threshold less deep in the tails of the distribution
than in the CW search,
namely at the mean plus one standard deviation of the propagated Gumbel distribution.
This same threshold is used
both for candidate identification
and later in the \uls procedure.
The results from the \numglitchsearchesfordistromax transient searches where we apply this method
are generally relatively clean,
as indicated by the batchmax histograms and goodness of the Gumbel distribution fits,
and so we do not employ any notching of disturbances across the board.
However, for glitches 7 and 8 of \ohfivethirtyseven additional features appear in the batchmax histograms and degrade the Gumbel fit.
In both cases, with three iterations of the automated notching procedure from~\citet{Tenorio:2021wad} we can produce clean batchmax histograms.
For glitch 7, the threshold on $\tCWdetstat$ is quite robust under notching
(changing from 11.3 to 11.0),
and no outliers are recovered.
On the other hand, for glitch 8,
the threshold is lowered sufficiently
(from 12.5 to 10.5)
which reclassifies the largest values at two frequencies as marginal outliers.
Follow-up of these outliers is detailed in Appendix~\ref{app:J0537-outliers}.

For \elevenohfive and \eighteentwentysix the number of templates is too small
to obtain robust Gumbel fits with the \dmax method.
Instead, we use the off-sourcing approach~\citep{Isi:2020uxj}:
we indirectly estimate the background distribution
by rerunning the full search
1000 times on different sky positions
but with otherwise identical settings.
This is feasible in these cases as the two searches are very cheap, less than a single core hour per run.
By off-sourcing on the sky we sample combinations of the same data that are statistically
independent from the templates used in the main search.
The practical implementation here matches that of~\citet{Tenorio:2021njf}:
we keep declination fixed and change only right ascension $\alpha$,
excluding a part of the sky closer than $0.5\pi$ to the pulsar
to ensure statistical independence.
We then fit a Gumbel distribution
to the set of the highest $\tCWdetstat$ values from each off-sourced search
and again set the threshold at the mean plus one standard deviation
of this distribution.

\section{Follow-up of remaining outliers for CW searches}
\label{app:final_follow_up}

In this section, we discuss follow-up of outliers found when searching in the direction of \psr{1838--0655} and \psr{1813--1749} (before glitch) with the \Fstat.
The same region of parameter space that produced outliers for \psr{1838--0655} in the \Fstat search, also produced outliers in the 5$n$-vector search.

\begin{figure}
    \centering
    \includegraphics[width=0.6\textwidth]{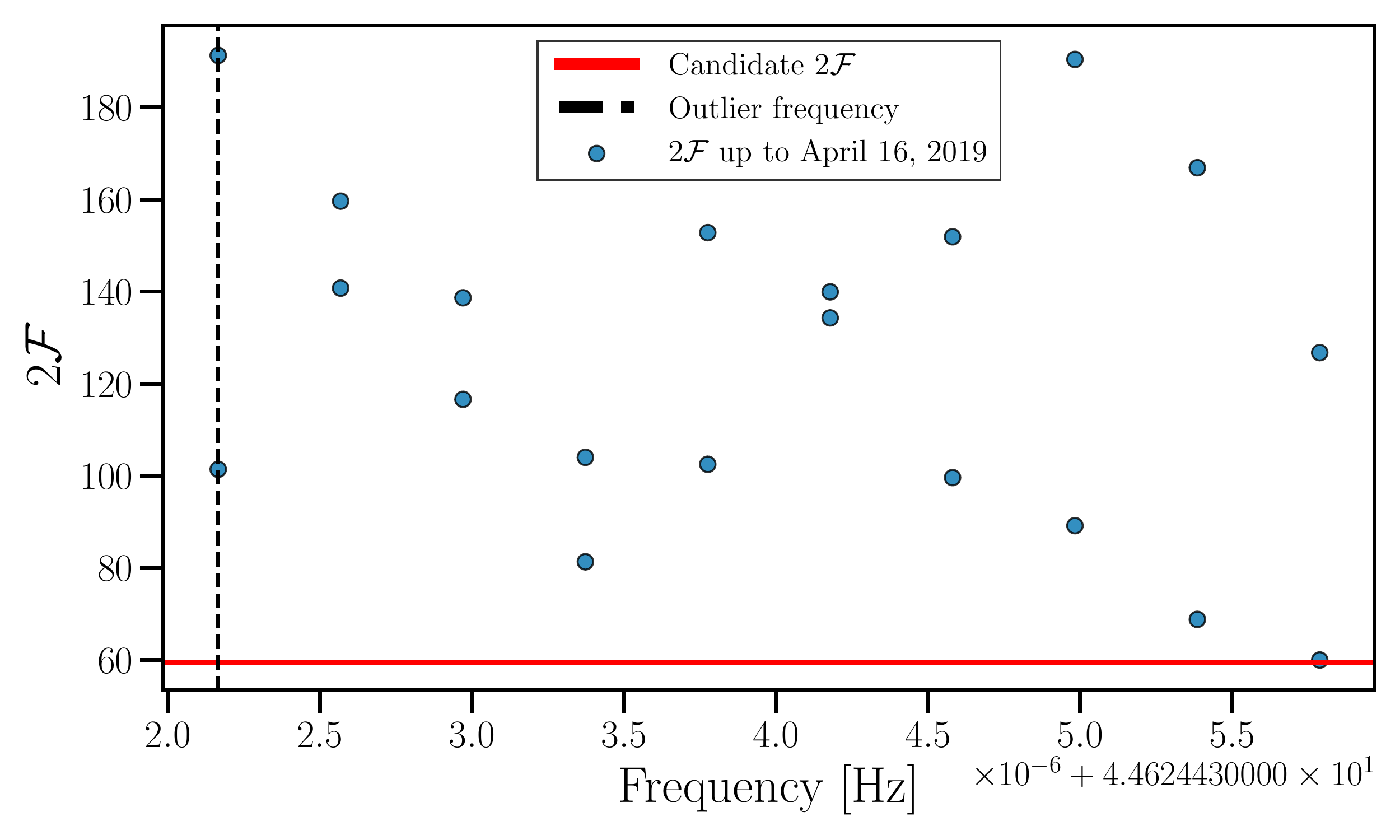}
    \caption{
    \label{fig:J1813_outlier_follow_up}
    $2\F$ values vs. frequency (blue circles) for a search from 2019 April 1 -- 2019 April 16 around the 44.62443216~Hz outlier in the \Fstat search for \psr{1813--1749} bg.
    The red line indicates the $2\F$ value associated with the outlier, while the vertical black dashed line shows its frequency.
    It is clear that the search on just 2 weeks of data produces larger $2\F$ values than the search on the full year of data, which is inconsistent with a true signal. The data for this figure are available online~\citep{data_release}.
    }
\end{figure}

The \Fstat results of \psr{1813--1749},
searching before the potential glitch,
show an outlier at 44.62443216~Hz,
close to a known line at H1 (see Fig.~\ref{fig:fstat_J1813-1749_results}).
However, the outlier frequency is outside the nominal width of the line specified in~\citet{linelist01}, and is therefore unlikely to be caused by that specific artifact.
On April 16, 2019, the frequencies of the calibration lines were changed to improve detector calibration which altered the character of the persistent narrowband artifacts seen in H1 data~\citep{Sun:2020wke}.
The known line in Fig.~\ref{fig:fstat_J1813-1749_results} is an example of one such artifact. To test whether this candidate is caused by a separate, but similar artifact, we run the search from the start of the run until only April 16, 2019.
In Fig.~\ref{fig:J1813_outlier_follow_up} we show that, zooming in on the frequency range around the outliers, 
quite a few templates have significantly higher values of $2\F$ than the outliers do for the full run. This behavior is inconsistent with a \cw signal, and so we veto these outliers.

\begin{figure}
    \centering
    \includegraphics[width=0.6\textwidth]{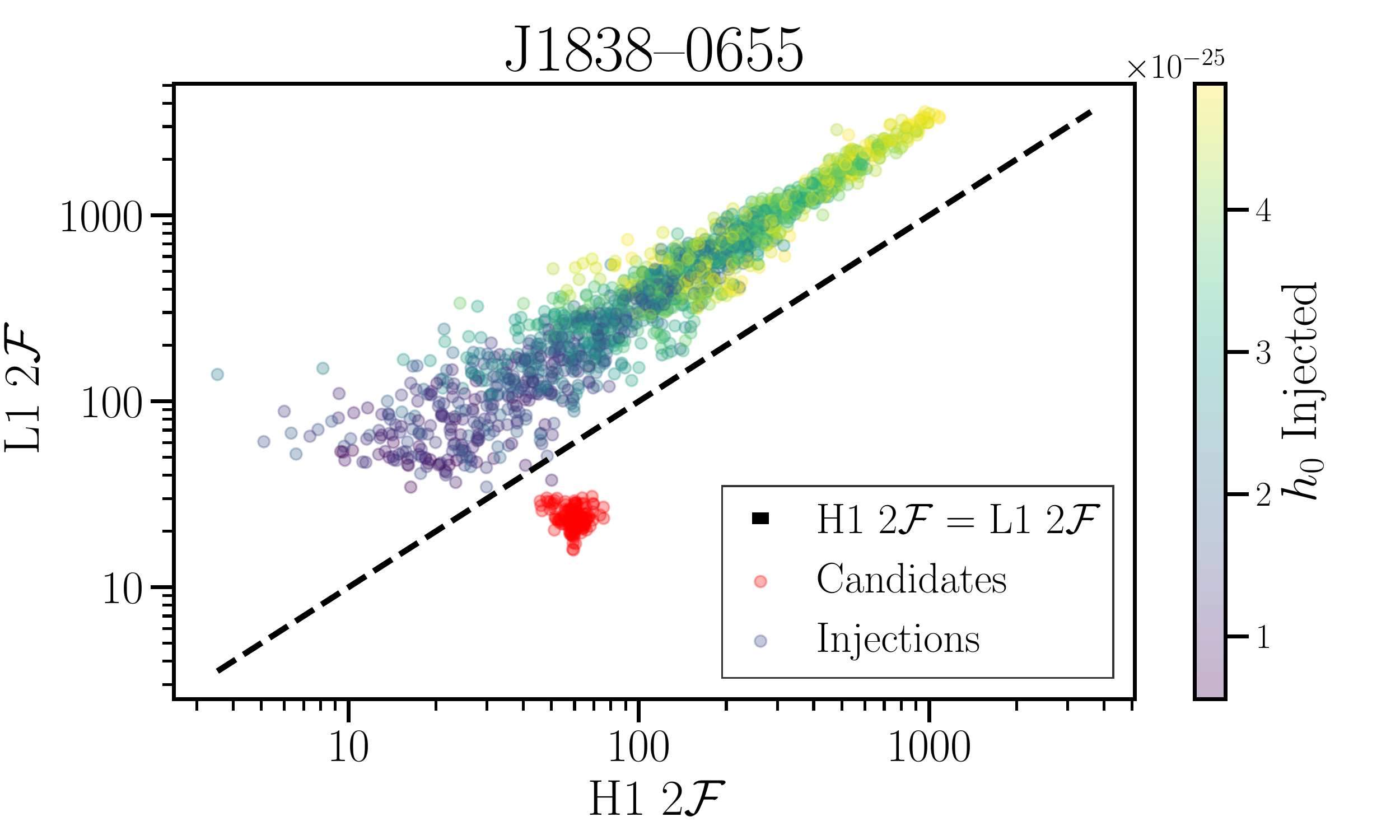}
    \caption{
    \label{fig:J1838_injection_results}
    Individual detector $2\F$ values calculated for software injections detected when calculating \uls (blue to yellow dots indicating low to high $h_0$ injections), and outliers (red dots).
    The dashed line indicates where $2\F$ is equal for the two detectors.
    We see the red points clustered away from the software injections.
    Only two of the software injection recoveries fall on the right hand side of the red line.
    The data for this figure are available online~\citep{data_release}.
    }
\end{figure}

In Fig.~\ref{fig:J1838m0655_S_candidates} we show that the outliers associated with \psr{1838--0655} are near an unknown disturbance at H1.
The $2\F$ values running only on H1 data are larger for all of these candidates than they are when running on only L1 data, despite L1 being more sensitive in this frequency band.
To test whether this is reasonable for a signal, we use the 1454 software injections that were detected above the search threshold when calculating \uls.
We plot the $2\F$ calculated using L1 on the y-axis and $2\F$ calculated using H1 on the x-axis of Fig.~\ref{fig:J1838_injection_results} for all injections (blue to yellow varied colors), and for the outliers (red).
The dashed line indicates where these two values are equal.
Only 2 out of 1454 injections cross the red line, meaning that if we veto any candidates to the right of the red line, we incur a false dismissal of 0.1\%.
Therefore, we veto this outlier in both the \Fstat and the 5$n$-vector searches.

\section{Follow-up of \ohfivethirtyseven glitch 8 transient search outliers}
\label{app:J0537-outliers}

Here we discuss follow-up analyses of the marginal outliers
recovered for the post-glitch transient search targeting
glitch 8 of the Big Glitcher \ohfivethirtyseven.
The parameters of the two outliers of interest are listed in Table~\ref{tab:J0537outliers}.
Another template with $\Btrans$ marginally above threshold is not offset by more than a single bin in any parameter from outlier A;
we therefore do not follow that one up separately.

\begin{deluxetable}{ccccccccc}
\tablecaption{
 \label{tab:J0537outliers}
 Outliers from the transient search after glitch 8 of \ohfivethirtyseven.
}
\tablehead{ & \colhead{$f$ [Hz]} & \colhead{$\fdot$ [$10^{-10}$\,Hz/s]} & \colhead{$\fddot$ [$10^{-20}$\,Hz/s$^2$]} & \colhead{$\tzero^\mathrm{ML}$ [s]} & \colhead{$\tau^\mathrm{ML}$ [s]} & \colhead{$\max2\F$} & \colhead{$\log_{10}\Btrans$} & \colhead{tail probability}}
\startdata
   A & 123.7934283 & -3.997 & -8.39 & 1263844981 & 5209200 & 53.4 & 11.00 & 4\%\\
   B & 123.8640925 & -3.995 & -1.89 & 1263628981 & 3913200 & 53.2 & 10.86 & 14\% \\
\enddata
\tablecomments{
The transient parameter estimates $\tzero^\mathrm{ML}$ and $\tau^\mathrm{ML}$ correspond to the location of the $\max2\F$ value in the $\Fmn$ map.
The posterior estimates (assuming flat priors, as for the $\Btrans$ statistic) are within 1 hour of the ML values,
with the exception of $\tau$ for outlier A
which is about 7 hours longer than the ML value.
The tail probability refers to the propagated Gumbel distribution
following the \dmax method
with three notching iterations.
}
\end{deluxetable}

First, we checked for any evidence that the outliers may be caused by instrumental artifacts.
There are no spectral artifacts of identified or unidentified origin listed in the relevant frequency band in \citet{linelist01} for LIGO nor \citet{virgo3lines} for Virgo.
Comparing the time-frequency tracks of the outliers with single-detector spectrograms~\citep{2019APS..APRY10007W} also reveals no suspicious structures.
As a multi-detector consistency check~\citep{Keitel:2013wga,Leaci:2015iuc},
we recomputed the $\Fmn$ maps at each outlier's $\Dop$ points for each individual detector.
The corresponding $\maxF$ and $\Btrans$ values and best-fit $(\tzero,\tau)$ pairs are collected in Table~\ref{tab:J0537outliersSingleIFO}.
Virgo does not recover these candidates,
as expected from its much lower sensitivity in this frequency range,
and returns unrelated short-duration peaks instead.
But the H1 and L1 results are fully consistent with each other
and with the multi-detector result.

\begin{deluxetable}{lcrrrrr}
\tablecaption{
 \label{tab:J0537outliersSingleIFO}
 Detector consistency follow-up for \ohfivethirtyseven glitch 8 outliers.
}
\tablehead{ \colhead{detector(s)} & \colhead{$\tzero^\mathrm{ML}$ [s]} & \colhead{$\tau^\mathrm{ML}$ [s]} & \colhead{$\max2\F$} & \colhead{$\log_{10}\Btrans$} & \colhead{$\tzero^\mathrm{MP}$ [s]} & \colhead{$\tau^\mathrm{MP}$ [s]}}
\startdata
outlier A & & & & & & \\
HLV & 1263844981 & 5209200 & 53.4 & 11.00 & 1263843607 & 5182844 \\
H1  & 1263758883 & 5284800 & 29.0 &  5.14 & 1263758773 & 5217811 \\
L1  & 1264317301 & 4669200 & 32.1 &  5.62 & 1263952957 & 5052102 \\
V1  & 1252638018 &   57600 & 27.5 &  3.05 & 1240493379 & 282595  \\
\tableline
outlier B & & & & & & \\
HLV & 1263628981 & 3913200 & 53.2 & 10.86 & 1263628503 & 3912769 \\
H1  & 1263629283 & 2196000 & 31.2 &  4.80 & 1263629183 & 2197636 \\
L1  & 1263590101 & 3945600 & 33.6 &  5.27 & 1263589393 & 4108995 \\
V1  & 1240437618 &    3600 & 28.3 &  2.95 & 1247483923 &    5400  \\
\enddata
\tablecomments{
These values are all obtained from the $\Fmn$ maps evaluated at the $\Dop$ parameters as listed for each outlier in Table~\ref{tab:J0537outliers}.
The transient parameter estimates $\tzero^\mathrm{ML}$ and $\tau^\mathrm{ML}$ correspond to the location of the $\max2\F$ value,
while $\tzero^\mathrm{MP}$ and $\tau^\mathrm{MP}$ are maximum-posterior estimates.
}
\end{deluxetable}

Other possible vetoes against an astrophysical origin include
checking whether or not the detection statistics would increase
when turning off Doppler modulation inside the \Fstat code~\citep[``DMoff'';][]{Zhu:2017ujz},
or searching at a different sky position~\citep[off-sourcing,][]{Isi:2020uxj}.
With a DMoff re-run of the full search band,
we recover neither of the outliers as a local maximum,
and the overall maximum at $\log_{10}\Btrans=10.1143$ is lower,
meaning that the outliers pass this check.
For off-sourcing, we perform 1000 analyses at different $\alpha$
(at least $0.5\pi$ away from \ohfivethirtyseven)
over 5\,mHz wide sub-bands of the original search.
This does not turn out to veto the outliers either,
as we do not find extended sets of similar or larger $\Btrans$ values
at many other sky positions and similar frequencies,
as could be expected from some types of instrumental artifacts.
However, we can once again fit a Gumbel distribution to the maxima
of these off-sourced searches
and propagate to the corresponding expected distribution
over the whole search bank
taking into account the missing trials factor of 25.
This results in tail probabilities of 2\% and 3\% for the two outliers,
which are lower than the 4\% and 14\% from \dmax
but still above the 1\% threshold
used by the two \cw search pipelines.

We also followed up both outliers with four separate pipelines:
(i) grid-less follow-up with \pyfstat~\citep{Ashton:2018ure,pyfstat,Keitel:2021xeq,Tenorio:2021njf};
(ii) the \Fstat \cw pipeline using \weave, as described in Sec.~\ref{sec:fstat-method};
(iii) the $5n$-vector method as described in Sec.~\ref{sec:fivevec-method};
and (iv) the targeted time-domain Bayesian method~\citep{Dupuis:2005,2017arXiv170508978P},
as also used for the targeted searches in \citet{O3KnownPulsars}.
In (i)
we use the same implementation of the transient \Fstat as in the main search.
But for (ii)--(iv),
we treat the candidates as putative \cw signals of fixed length,
fixing the start time $\tzero$ and end time $\tzero+\tau$
of these analyses
based on the outlier parameters from Table~\ref{tab:J0537outliers}.

The \pyfstat package~\citep{pyfstat,Keitel:2021xeq} contains an implementation
of Markov-Chain Monte Carlo (MCMC) grid-less follow-up~\citep{Ashton:2018ure,Tenorio:2021njf}
using the \code{ptemcee} sampler~\citep{Foreman-Mackey:2012any,Vousden:2015pte}
and the same \Fstat code in \lalsuite~\citep{LALSuite,Wette:2020air}.
We do not employ this here as a veto or for significance estimation,
but only to check that the candidates can be recovered independently of the original search grid setup,
when putting Gaussian priors on the $\Dop$ parameters
around their previously recovered values,
and now also allowing variations in sky position.
These \pyfstat analyses recover consistent $\maxF$ values and locations.

The \weave pipeline also uses the same underlying \Fstat implementation.
The main difference is to not search over different signal start times and durations,
instead fixing start time $\tzero$ and end time $\tzero+\tau$
and only searching over $\Dop$ parameters.
As expected, this follow-up recovers \Fstat peaks consistent with the outlier parameters.

With the $5n$-vector method
and a similar follow-up setup as for \weave,
we again recover local peaks at the frequency of each outlier.
But over the given data span,
this follow-up also recovers other peaks with higher $S$-statistic values
across the probed 0.45\,Hz bands and spin-down ranges.
Thus, it does not confirm the outliers as candidates of interest.
For outlier A, there are 15 peaks with higher detection statistic
(corresponding to 0.36\% of the top level candidates in every $10^{-4}$~ Hz sub-band and marginalized over the spin-down region).
For outlier B, there are 5 peaks with higher detection statistic (0.15\% of the top level candidates in every $10^{-4}$~ Hz sub-band and spin-down region).

Finally, the targeted time-domain Bayesian method fixes the signal duration as well as the $\Dop$ parameters,
and performs nested sampling over the four amplitude parameters $\Amp$.
For both outliers, this returned Bayes factors of about $500$ in favor of a coherent signal in all three detectors as opposed to noise or incoherent signals in the individual detectors.
However, this fully targeted analysis does not provide any additional assessment of statistical significance.
In addition, the estimated strain amplitudes $h_0$ from this follow-up,
$4.5^{+1.5}_{-1.4}\times10^{-26}$ and $5.2^{+1.7}_{-1.7}\times10^{-26}$ at 95\% credible intervals,
are a factor 10 above the indirect glitch energy \ul from Eq.~\ref{eq:glitchenergyUL}
assuming a signal from \ohfivethirtyseven and with the NICER-observed glitch size.

\allauthors
\end{document}